\documentclass[aps,prd,reprint,amsmath,amssymb]{revtex4-2}
\usepackage{url}
\usepackage{color}
\usepackage{graphicx}
\usepackage{overpic}
\usepackage{dcolumn}
\usepackage{bbm}
\usepackage{bm}
\usepackage{subfigure}
\usepackage{amsfonts}
\usepackage{amsmath}
\usepackage{amssymb}
\usepackage{sidecap}
\usepackage{float}
\usepackage{parskip}
\usepackage{grffile}

\usepackage{hyperref}
\usepackage{hhline}
\usepackage{mathtools}
\usepackage{comment}
\usepackage[normalem]{ulem}

\makeatletter
\def\@eqnnum{{\normalsize \normalcolor (\theequation)}}
 \makeatother
\begin{document}

\title{Multiple hysteresis widths in inertial Kuramoto model}
\author{Jayesh C. Jain and Sarika Jalan}
\email{Corresponding Author: sarikajalan9@gmail.com} \affiliation{Complex Systems Lab, Department of Physics, Indian Institute of Technology Indore, Khandwa Road, Simrol, Indore-453552, India}

\begin{abstract}
Multistability is a well-known feature of the inertial Kuramoto system (KMI). Here, we demonstrate that an interplay of phase lag and triadic interactions in KMI leads to distinct hysteresis widths corresponding to different stable states. This phenomenon becomes more pronounced with increasing inertia. Theoretical calculations for the backward branch based on self-consistent analysis show that these multiple widths arise from saddle-node bifurcation occurring at different coupling strengths. Moreover, the forward branch corresponds to oscillatory state and does not admit steady-state solution.
The study of multiple hysteresis widths may be useful in modeling power grid systems, information storage, and memory selection in real-word systems.
\end{abstract}

\maketitle
\section{Introduction}
A broad class of complex systems observed in nature and studied across diverse fields, such as physics, neuroscience, chemistry, and biology, can be described using coupled oscillators \cite{Winfree_book,strogatz_book,Strogatz20001,osipov2007}. Synchronization refers to the process by which a collection of oscillators adjusts its rhythms through mutual interactions, ultimately oscillating at a common frequency even though each oscillator has its own intrinsic frequency. Winfree \cite{Winfree1967} and Kuramoto \cite{kuramoto01} introduced simple models to capture this synchronization phenomenon. Kuramoto model, in particular, has become a standard framework for analyzing synchronization. In this model, oscillators evolve according to their individual intrinsic frequencies, and effect of one oscillator on another is usually represented by a sinusoidal coupling between their phases. The Kuramoto model has been widely employed to describe collective dynamics in systems ranging from chorusing birds to neuronal firing patterns in the brain  \cite{Strogatz20001,strogatz_book,acebron2005}. 

Since its formulation, many works have explored various generalizations and extensions of the Kuramoto model \cite{networks,acebron2005,RODRIGUES2016,pikovsky_review}. Tanaka \textit{et al.} \cite{tanaka1997,tanaka-prl} proposed a key modification to the standard Kuramoto model by including an inertial term to originally describe the behavior of certain types of fireflies (\textit{Pteroptyx malaccae}) \cite{Ermentrout}. Later, Kuramoto model with inertia (KMI) has proven to be useful for describing Josephson junction arrays \cite{levi1978,WATANABE1994,stroud2005}, goods markets \cite{ikeda2011}, dendritic neurons \cite{SAKYTE2011}, and power grids \cite{acebron2005,nielsen2008,review}. With the addition of inertia, Kuramoto oscillators display richer and more complex dynamical behaviors, notably, first-order (explosive) transitions to synchronization, hysteresis, bistability, and multistability \cite{tanaka1997,olmi2015,gao2018,p.ji2013}. Olmi \textit{et al.} \cite{olmi2015} reported that multiple partially synchronized states of various sizes coexist (coexistence of multiple attractors) at a given parameter value referred as the multistability of the KMI. Specifically, the occurrence of hysteresis in the KMI is due to the coexistence of fixed points and limit cycles. Tanaka \textit{et al.} \cite{tanaka1997} and Olmi \textit{et al.} \cite{olmi2015} described hysteretic behavior by extending the self-consistent mean-field approach for KMI. More recently, Gao and Efstathiou \cite{gao2021} highlighted the presence of oscillatory and standing wave states in KMI for unimodal and bimodal frequency distributions. 

The role of phase lag in Kuramoto oscillator \cite{sakaguchi1986} (Kuramoto-Sakaguchi) with identical frequencies including inertia has been emphasized in several works, indicating the presence of chimeras \cite{olmi-chimera, MUNYAYEV2024}, cyclops states \cite{cyclops}, and frequency clustering \cite{ashwin,krischer}. For non-identical frequencies, Barr\'e and M\'etiviere \cite{barre} analytically indicated that even small inertia can change the nature of synchronization transitions in Kuramoto models by performing the method of unstable manifold expansion \cite{crawford1, crawford2}. Gao and Efstathiou \cite{gao2020} discussed the self consistent approach for the KMI model with phase lag for pairwise interactions, and demonstrated the existence of partially synchronized states due to interplay between inertia and phase lag.

Furthermore, inclusion of higher-order interactions which goes beyond transitional pair-wise couplings has been shown to provide a better model to characterize functionality of many real-world complex systems \cite{ cyclops, skardal, xu}. Incorporating higher-order interactions in the Kuramoto model has been shown to lead abrupt (de)synchronization \cite{skardal} and tiered synchronization \cite{priyanka1}. Incorporating higher-order coupling in KMI, Sabhahit et al. \cite{narayan2024} signaled the presence of prolonged hysteresis in its parameter space. While, the effects of triadic interactions on KMI dynamics have been studied without phase lag, the effects of phase lag have been studied in KMI without higher-order interactions only. The combined effect of higher-order and phase lag still remains unknown for KMI. 
\begin{figure*}[ht]
    \centering
    \subfigure{
        \includegraphics[width=0.4\textwidth]{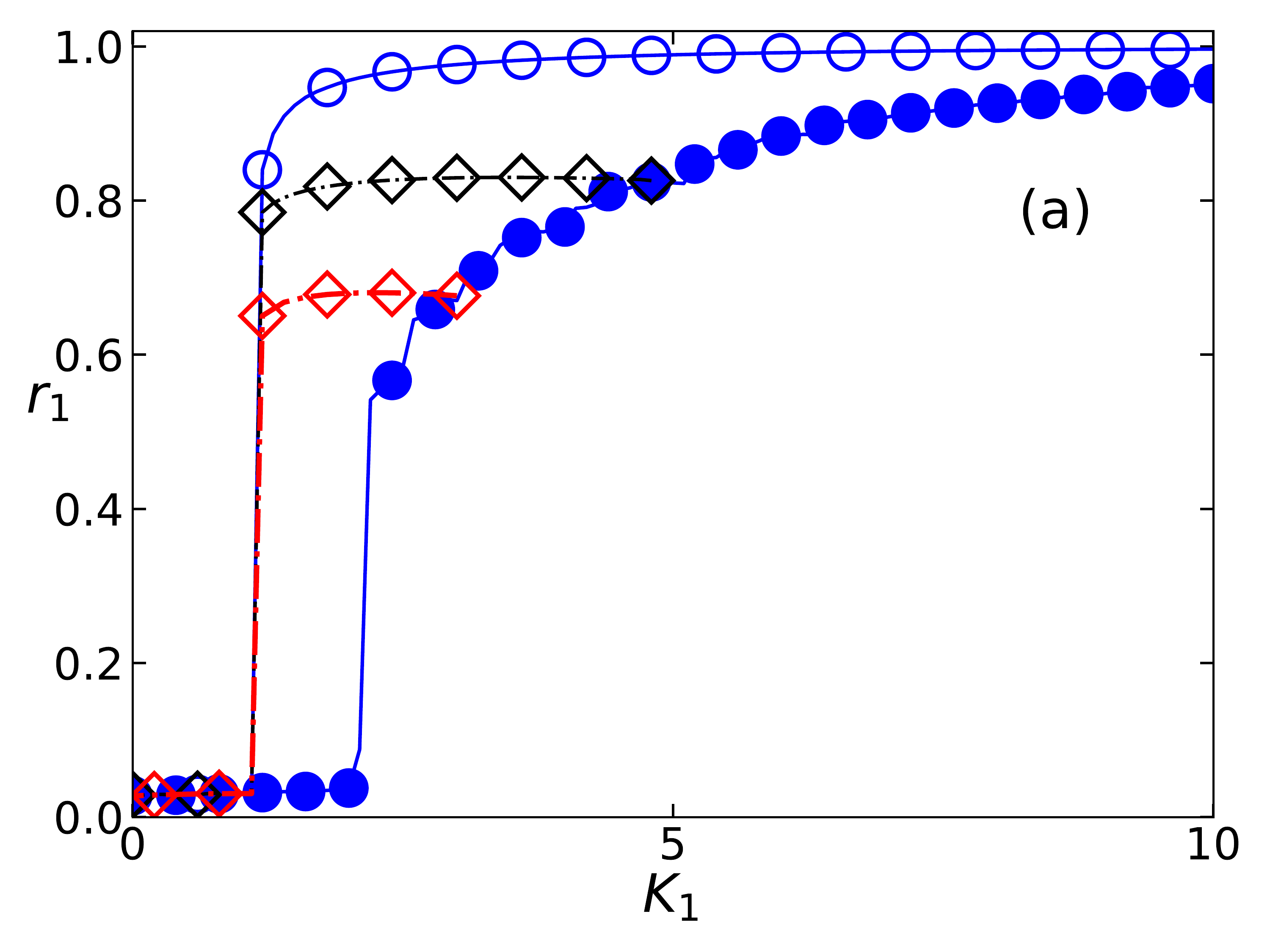}
        \label{multi-1}
    }
    \hspace{1cm}
    \subfigure{
        \includegraphics[width=0.4\textwidth]{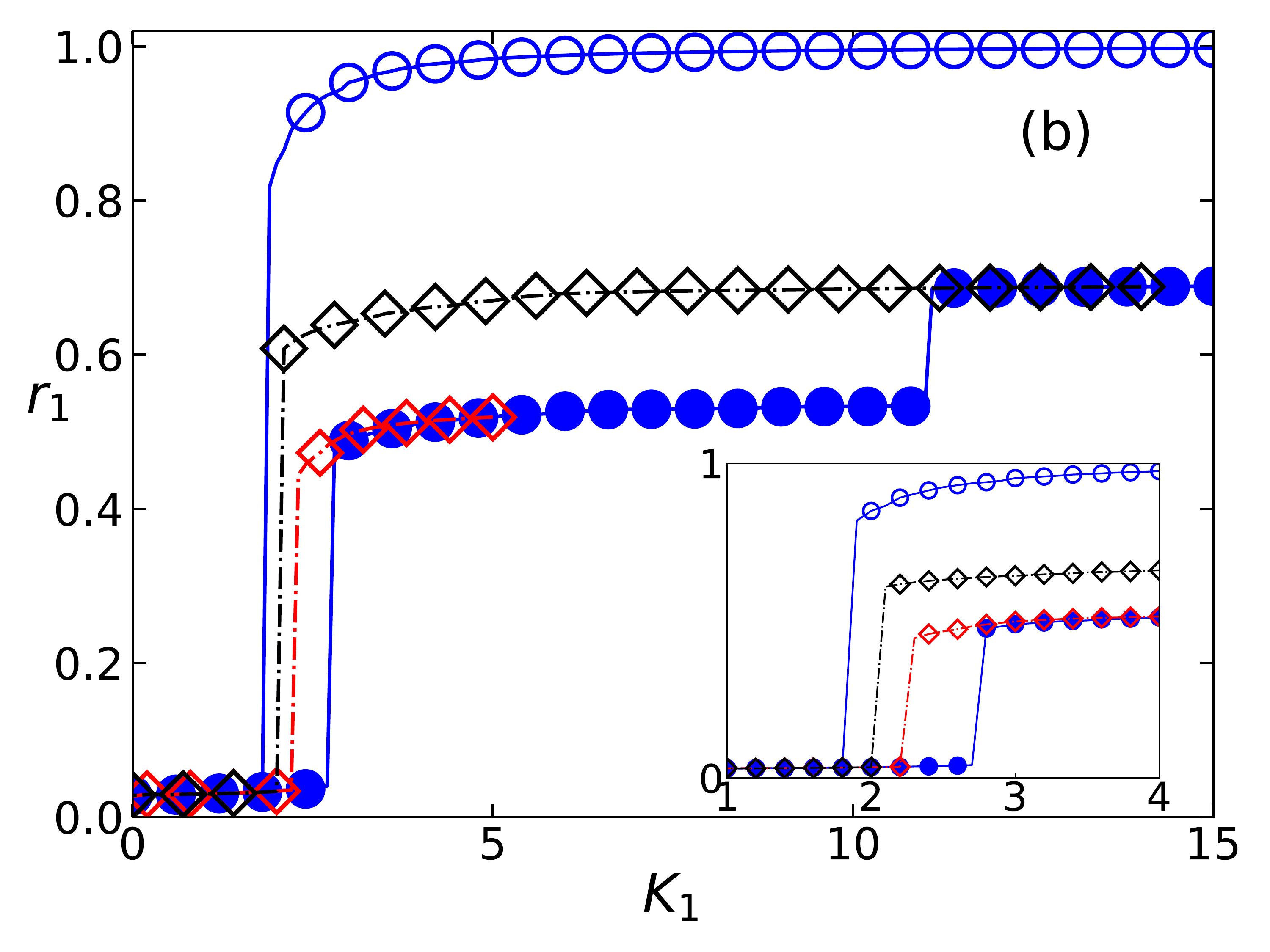}
        \label{multi-2}
    }
    \caption{(Color online) Multiple hysteresis in KMI with triadic interactions arising due to phase-lag.  $r_1$ is plotted as a function of $K_1$ (Eq.~\ref{model-meanfield}). (a) $\alpha=0$: A single hysteresis width is observed.
    (b) $\alpha = \pi/6$: Distinct hysteresis widths emerges depending upon the initial state. The solid (hollow) blue circles denote the forward (backward) process, and hollow diamonds with dash-dotted lines (red and black) denote backward branches from two different initial conditions corresponding two different branches. Inset is zoomed version to clearly demonstrate initial state dependent hysteresis widths. Here, $m=4$, $K_2 = 2$, and $N=1000$.}
    \label{fig:multistable}
\end{figure*}

Filling this gap, here we investigate the KMI with phase lag and triadic interactions. Standard dimensional-reduction techniques, such as the Ott-Antonsen (OA) approach \cite{OA}, do not apply to KMI models, as the density function explicitly depends on the velocity. Hence, we use the self-consistency method \cite{tanaka-prl} to obtain analytical solutions. Here, coupled KMI is characterized by forward and backward transitions. In the forward branch, starting from an incoherent state, the coupling strength increases. Whereas, the backward branch is obtained by starting the simulation from a coherent synchronized state, and decreasing the coupling strength adiabatically.
The primary challenge lies in analyzing the parameter space due to the presence of phase lag. To overcome this, we transform the variables in auxiliary space to obtain the analytical solution. We further note that the forward branch does not admit steady-state solutions and is time dependent and oscillatory. 

The paper is organized as follows: Sec.~\ref{model_sys} describes the model and represents the closed form of the solution in terms of new variables. Sec.~\ref{numerics} presents the results from numerical simulation of the model equation. Sec.~\ref{analytics} shows the analytical results for the steady-state solution, and presents the influence of phase lag, inertia, and higher-order interactions on synchronization. Sec.~\ref{two-cluster} provides the reason behind the partially synchronized behavior of the model within the framework of two-cluster synchronization. We conclude the study in Sec.~\ref{conclusion}.

\section{The Model}\label{model_sys}
We consider a system of $N$ globally coupled oscillators governed by the second order Kuramoto-Sakaguchi model \cite{sakaguchi1986}. The dynamics of $i^{\text{th}}$ oscillator can be written as:
\begin{multline}\label{model}
    m \ddot{\theta_i} + \dot{\theta_i} = \omega_i + \frac{K_1}{N} \sum_{j=1}^N \sin(\theta_j - \theta_i - \alpha) \\+ \frac{K_2}{N^2}\sum_{j,k=1}^N \sin(2\theta_j - \theta_k - \theta_i - \alpha).
\end{multline}
Here, $i=1,2,...N$ while the term $m$ represents inertia, $K_1$ and $K_2$ are coupling strengths of pairwise and triadic interactions, respectively. The natural frequency of $i^{\text{th}}$ oscillator is denoted as $\omega_i$ which follows the frequency distribution $g(\omega)$, and $\alpha$ is a constant phase lag. The dynamics of Eq.~(\ref{model}) can be simplified in terms of mean field defined as
\begin{equation} \label{meanfield}
    z_p = r_p e^{i\psi_p} = \frac{1}{N}\sum_{j=1}^N e^{i p\theta_j}  \hspace{0.5cm} ; \hspace{0.5cm} p=1,2.
\end{equation}

Here, $r_1 \in [0,1]$ provides a degree of global coherence of the system, while $r_2$ measures the 2-clusters state. The value $r_1=1$ indicates that the oscillators are synchronized with the same phases $\theta_j(t) = \theta(t)$ for all $j=1,2,..N$. On the other hand, if all oscillator phases($e^{i \theta_j}$) are uniformly distributed among the unit circle, $r_1\approx 0$. However, the converse is not always true as $r_1 \approx 0$ may arise due to symmetric phase clusters In those cases, the second order parameter $r_2$ becomes important as it quantifies the presence of 2-clusters in the system. The mean field described by Eq.~(\ref{meanfield}) reduces Eq.~(\ref{model}) as follows:
\begin{multline}\label{model-meanfield}
    m \ddot{\theta_i} + \dot{\theta_i} = \omega_i + K_1 r_1 \sin(\psi_1 - \theta_i - \alpha) \\+ K_2 r_2 r_1 \sin(\psi_2-\psi_1-\theta_i -\alpha).
\end{multline}
To gain further insight into dynamics, we transform the variables in a rotating frame given by: $\psi_1 = \Omega t + \nu$ and take $\theta \rightarrow \theta + \Omega t+\nu$. Hence, Eq.~(\ref{model-meanfield}) in a rotating frame is written as
\begin{multline*}
    m \ddot{\theta_i} + \dot{\theta_i} = \omega_i - \Omega - K_1 r_1 \sin(\theta_i + \alpha) \\- K_2 r_2 r_1 \sin(\theta_i+\alpha + \gamma),
\end{multline*}
here, we have taken $\gamma = 2 \psi_1 - \psi_2$ and set $\nu = 0$ without loss of generality. 

To study the parameter space, we define new auxiliary variables as 
$R\cos{A} = K_1 r_1 + K_2 r_1 r_2 \cos{\gamma}$ and $R \sin{A}= K_2 r_1 r_2 \sin{\gamma}$. Dropping the subscript $i$, we get 
\begin{equation}\label{tt}
    m \ddot{\theta} + \dot{\theta} = (\omega - \Omega) - R \sin(\theta + \beta),
\end{equation}
where, the auxiliary parameters are defined as 
\begin{align*}
    R &= \sqrt{(K_1r_1)^2+(K_2r_2r_1)^2 + 2 K_1K_2(r_1)^2r_2 \cos{\gamma}};\\
    \beta &= \alpha + A =  \alpha +\arcsin{\left(\frac{K_2r_1r_2 \sin{\gamma}}{R}\right)}.
\end{align*}
To study the parameter space and the analytical solution, Eq.~(\ref{tt}) is important to obtain the limits of the steady-state solution.

\section{Numerical Results}\label{numerics}
We numerically evolve Eq.~(\ref{model}) using the fourth-order Runge-Kutta (RK-4) method with a step size $\delta t = 0.01$ for time $T=5400$. The natural frequency distribution is chosen as a Gaussian frequency distribution $g(\omega) = \frac{1}{\sqrt{2\pi}}e^{-\omega^2/2 \sigma^2}$ with zero mean and standard deviation $\sigma=1$. It is important to describe the initial conditions $(\theta(0),\dot{\theta}(0))$ used in the simulations, as they reveal the initial basin of attraction from which the system evolves. We perform simulation by varying $K_1$ adiabatically with two different procedures \cite{olmi2015} corresponding to two different initial conditions after discarding the transient part; procedure (I) depicts the phases $\theta_i$ are drawn randomly from the uniform distribution in $[0,2\pi]$ and the frequencies $\dot{\theta}_i$ from the uniform distribution in $[0,1]$. After that, we increase the value of $K_1$ by $\delta K_1 = 0.1$ adiabatically, where the initial condition of $K_1+\delta K_1$ is taken from the final state of $K_1$. procedure (II) refers to the simulation starting with all oscillators in the coherent state at $2\pi$. We start the simulation with a high value of $K_1$ and decrease $K_1$ by $\delta K_1$ adiabatically.
In KMI, the transition points obtained via procedures (I) and (II) generally do not coincide with each other, leading to hysteresis \cite{tanaka1997,tanaka-prl,olmi2015}.

To elucidate the role of higher-order interactions and phase lag, we compare two different cases [Fig.~(\ref{fig:multistable})]. It is evident that when both phase lag and triadic interactions are present, the hysteresis originating from the globally coherent branch differs from that arising from the multistable branch. The distinct path followed by procedures (I) and (II) arise due to the combined effect of $m$ and $K_2$. As illustrated in Fig.~(\ref{multi-1}), the values corresponding to forward and backward transition points are $(K_{1c})^f \approx 2$ and $(K_{1c})^b \approx 1.1$, respectively. While, the backward transition points for branches that initialized from two different multistable states are denoted as $\tilde{K}_{1c}^{(1)}$ and $\tilde{K}_{1c}^{(2)}$ and their numerical value is approximately $1.1$. 
To study the behavior qualitatively, we define a new variable called the ``hysteresis width" as $\Delta K_1 = |(K_{1c})^f -(K_{1c})^b|$. While, the backward transition point initiated from $p^{\text{th}}$ multistable branch is denoted by $\tilde{K}_{1c}^{(p)}$. The appearance of saddle-node bifurcation at different coupling strengths leads to different $\tilde{K}_{1c}$ values. The hysteresis width associated with a globally coherent state is denoted as $\Delta K_1$, while the hysteresis width corresponding to $p^{\text{th}}$ multistable branch is denoted as $\Delta \tilde{K}_{1}^{(p)}$. For non-zero phase lag (Fig.~(\ref{multi-2})), the transition points for forward and backward cases are $(K_{1c})^f \approx 2.7$ and $(K_{1c})^b \approx 1.8$, respectively, and the backward transition initiated from two different multistable states are $\tilde{K}_{1c}^{(1)} \approx 2$ and $\tilde{K}_{1c}^{(2)} \approx 2.2$, respectively. Hence, due to the interplay between phase lag and triadic term, the hysteresis widths associated with two multistable states ($\Delta \tilde{K}_1^{(1)}= 0.7$) and ($\Delta \tilde{K}_1^{(2)} = 0.5$) are not the same as those initiated from the globally coherent state ($\Delta K_1 = 0.9$). The associated hysteresis widths starting from different multistable states are smaller than the hysteresis width corresponding to procedure (II), i.e., $\Delta K_1>\Delta \tilde{K}_1^{(1)}>\Delta \tilde{K}_1^{(2)}$. The combined effect of phase lag and triadic interaction leads to shrinking of the hysteresis width, i.e., $\Delta K_1>\Delta \tilde{K}_{1}^{(p)}$ for a given multistable state. The width of these hysteresis varies depending on the initial multistable state. It should be noted that the maximum hysteresis width corresponding to a given multistable state is traced from the final stable state obtained by procedure (I), while the minimum hysteresis width is associated with the first stable state. We remark that the combined effect of inertia and phase lag leads to an oscillation of the order parameter in time, leading to partial synchronized behavior in the forward branch, as previously reported by Gao and Efstathiou.\cite{gao2020}.

\section{Analytics}\label{analytics}
To understand the dynamics associated with Eq.~(\ref{tt}), we consider the continuum case ($N \rightarrow \infty$) and perform a self-consistency analysis to obtain the values of $r_1$ and $r_2$ as function of $K_1$ for fixed values of $\alpha$ and $K_2$. The order parameter for $p \in \{1,2\} $ is defined as 
\begin{equation*}
    z_p = \int_{-\infty}^{\infty
   } \int_{-\pi}^{\pi} e^{i p \theta} f(\theta,\omega) g(\omega) d\theta d\omega, 
\end{equation*}
where, $f(\theta,\omega)d\theta$ is the probability density of the oscillator with their phases between $\theta$ to $\theta + d \theta$. The natural frequency distribution is taken to be Gaussian with zero mean and standard deviation $\sigma=1$.
\begin{equation*}
    g(\omega) = \frac{1}{\sqrt{2\pi}} e^{-(\omega^2/2\sigma^2)}.
\end{equation*}

In steady state, the order parameters $z_1$ and $z_2$ can be split into two parts depending on their intrinsic frequencies. One group of oscillators is locked at the mean phase, while the other oscillators drift over the entire circle. The order parameter associated with locked oscillators is denoted as $z^{(l)}_1$ and $z^{(l)}_2$, while for drifting oscillators it is defined as $z^{(d)}_1$ and $z^{(d)}_2$. Hence,
\begin{equation*}
    z_p = z^{(l)}_p+ z^{(d)}_p.
\end{equation*}
In terms of rescaled timescale, $t \rightarrow \tau \left(\sqrt{\frac{m}{R}}\right)$, we can rewrite Eq.~(\ref{tt}) as follows,
\begin{equation} \label{red_eq}
    \ddot{\theta} + \rho\dot{\theta} = \Gamma -\sin(\theta + \beta).
\end{equation}
Here, the parameters defined as $\rho = 1/\sqrt{mR}$ and $\Gamma = (\omega - \Omega)/R$ are dimensionless, and the dot represents the derivative with respect to $\tau$.

\begin{figure}[t]
    \centering
    \includegraphics[width=0.8\linewidth]{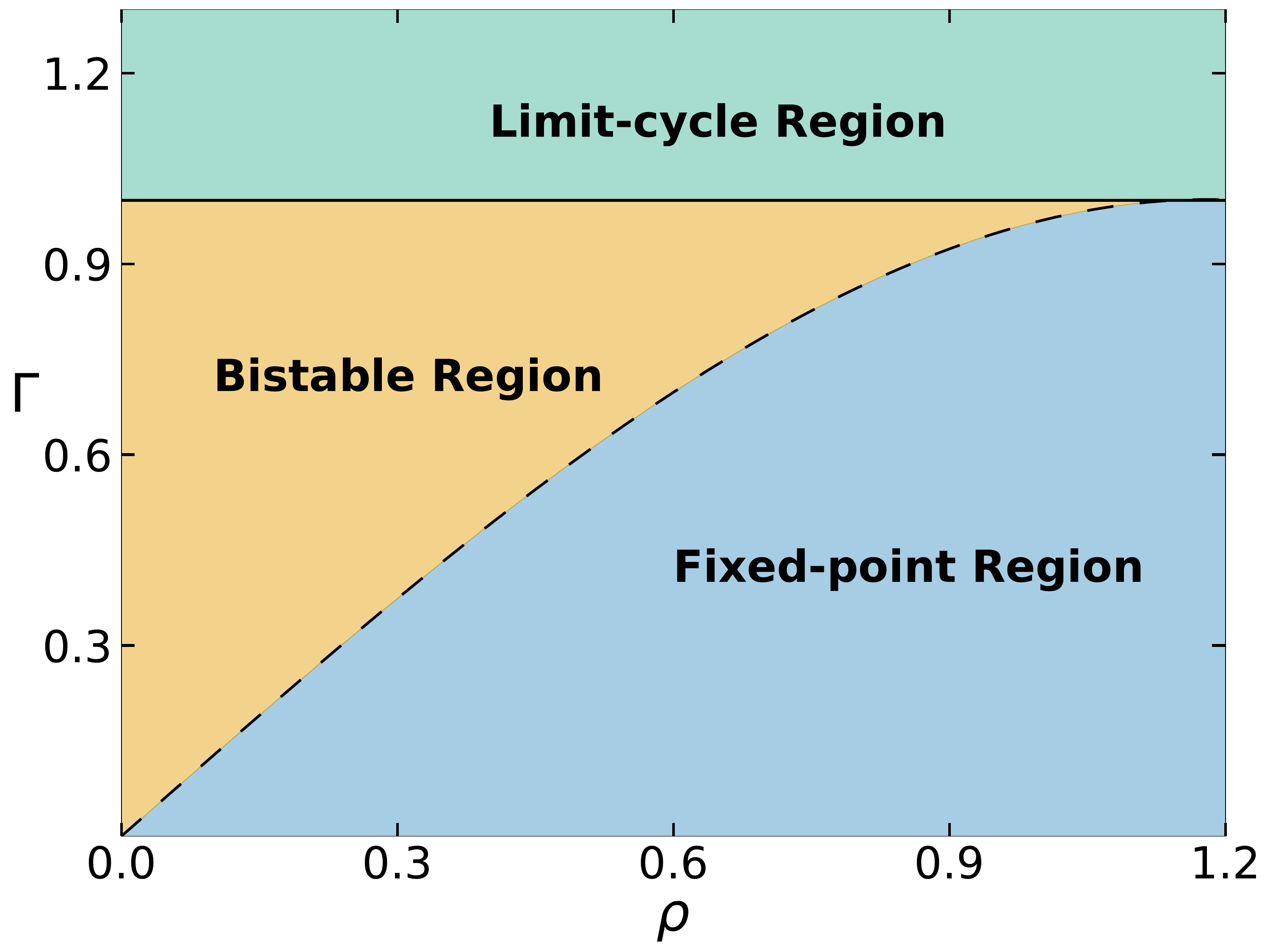}
    \caption{(Color online) $\rho-\Gamma$ parameter space obtained via Eq.~(\ref{red_eq}) consists of three different regions: Fixed point region, Limit cycle region and a Bistable region.}
    \label{fig:param_space}
\end{figure}

It has been shown\cite{narayan2024, gao2018, gao2020} that for a set of given parameter values, KMI consists of fixed points or a globally attracting limit cycle or a bistable region where both the limit cycle and a fixed point coexist. Setting $\ddot{\theta} = \dot{\theta} = 0$, Eq.~(\ref{red_eq}) yields two fixed point solutions for $\Gamma<1$, namely a stable node and a saddle. At $\Gamma=1$, these two fixed points coalesce and annihilate in a saddle-node bifurcation. For $\Gamma>1$, a stable limit cycle emerges in the phase space. Fig.~(\ref{fig:param_space}) indicates that this limit cycle disappears through a homoclinic bifurcation, which is well approximated by a straight line $\Gamma \approx (4/\pi) \rho$ predicted by the Melnikov method\cite{Melnikov}. 
Hysteresis in KMI is associated with the bistable regime  \cite{tanaka1997}. To quantify the dynamics, we treat forward and backward processes separately within the self-consistency framework. In the forward process, we initialize with a small value of $K_1$, where $r_1 \approx 0$ and the system is in the incoherent state. This corresponds to high values of $\Gamma$ and $\rho$, associate wtih the limit cycle region. As $K_1$ increases adiabatically, the system remains within the basin of attraction of the limit cycle until $\Gamma \leq (4/\pi)\rho$ is satisfied. However, for the backward process, we initialized at high $K_1$ value, where the oscillators are in the fixed point regime. As $K_1$ decreases adiabatically, the oscillators remain locked until $\Gamma=1$. Thus, in the forward process, oscillators having $|\omega -\Omega| \leq \frac{4}{\pi}\sqrt{\frac{R}{m}} = \omega_f $ contribute to the locked case, while in the backward process, oscillators having $|\omega - \Omega| \leq R = \omega_b$ are in locked states. The oscillators with $|\omega-\Omega| > \omega_{f,b} $ drift around the locked cluster. 

The contribution of locked oscillators ($z_p^{(l)}$) to coherence is calculated by, 
\begin{equation}
    z_p^{(l)} = \int _{-\omega_{f,b} +\Omega} ^{\omega_{f,b}+\Omega} e^{ip \theta^*} g(\omega) d\omega,
\end{equation}
where, fixed point is given by $\theta^* = \arcsin{(\Gamma)} - \beta$.
However, drifting part contributes to coherence as
\begin{equation*}
    z_p^{(d)} = \int_{|\omega-\Omega|> \omega_{f,b}} \int_{-\pi}^{\pi} e^{ip\theta} f_d(\theta,\omega) g(\omega) d\omega.
\end{equation*}

Here, $f_d(\theta,\omega)$ is the density of drfiting oscillators which satisfies $f_d \propto 1/|\dot{\theta}|$. The normalization condition $\int_{-\pi}^{\pi} f_d(\theta,\omega) d\theta = \int_0^Tf_d(\theta,\omega) \dot{\theta} dt =1$ yields the following
\begin{equation}\label{ww}
    z_p^{(d)} = \int_{|\omega-\Omega|> \omega_{f,b}} \left[\frac{1}{T} \int_0^Te^{ip\theta} dt \right] g(\omega) d\omega.
\end{equation}
Appendix~(\ref{app1}) contains the relevant details of analytical calculation of $z_1$ and $z_2$, and further numerically shows that in the backward process, $\psi_2 \approx 2 \psi_1$ yielding $\gamma \approx 0$. The reduced equations are then described by separating the real and imaginary parts of the solution $z_1$ as
\begin{multline}\label{a_r1}
    r_1  = \int_{-\infty}^{\infty}\Bigg[\mathbbm{1}_l(\sqrt{1-\Gamma^2}\cos{\beta} + \Gamma \sin{\beta}) + \\ \mathbbm{1}_d M \left(\frac{\Gamma}{\rho}\cos{\beta} - \rho\sin{\beta}\right) \Bigg] g(\omega) d\omega,
\end{multline}

\begin{multline}\label{a_psi1}
    0 = \int_{-\infty}^{\infty}\Bigg[\mathbbm{1}_l(\Gamma \cos{\beta} - \sqrt{1-\Gamma^2} \sin{\beta}) - \\ \mathbbm{1}_d M \left(\frac{\Gamma}{\rho}\sin{\beta}+ \rho \cos{\beta}\right) \Bigg]  g(\omega) d\omega,
\end{multline}
\begin{figure}[t]  
    \centering
        \includegraphics[width=0.4\textwidth]{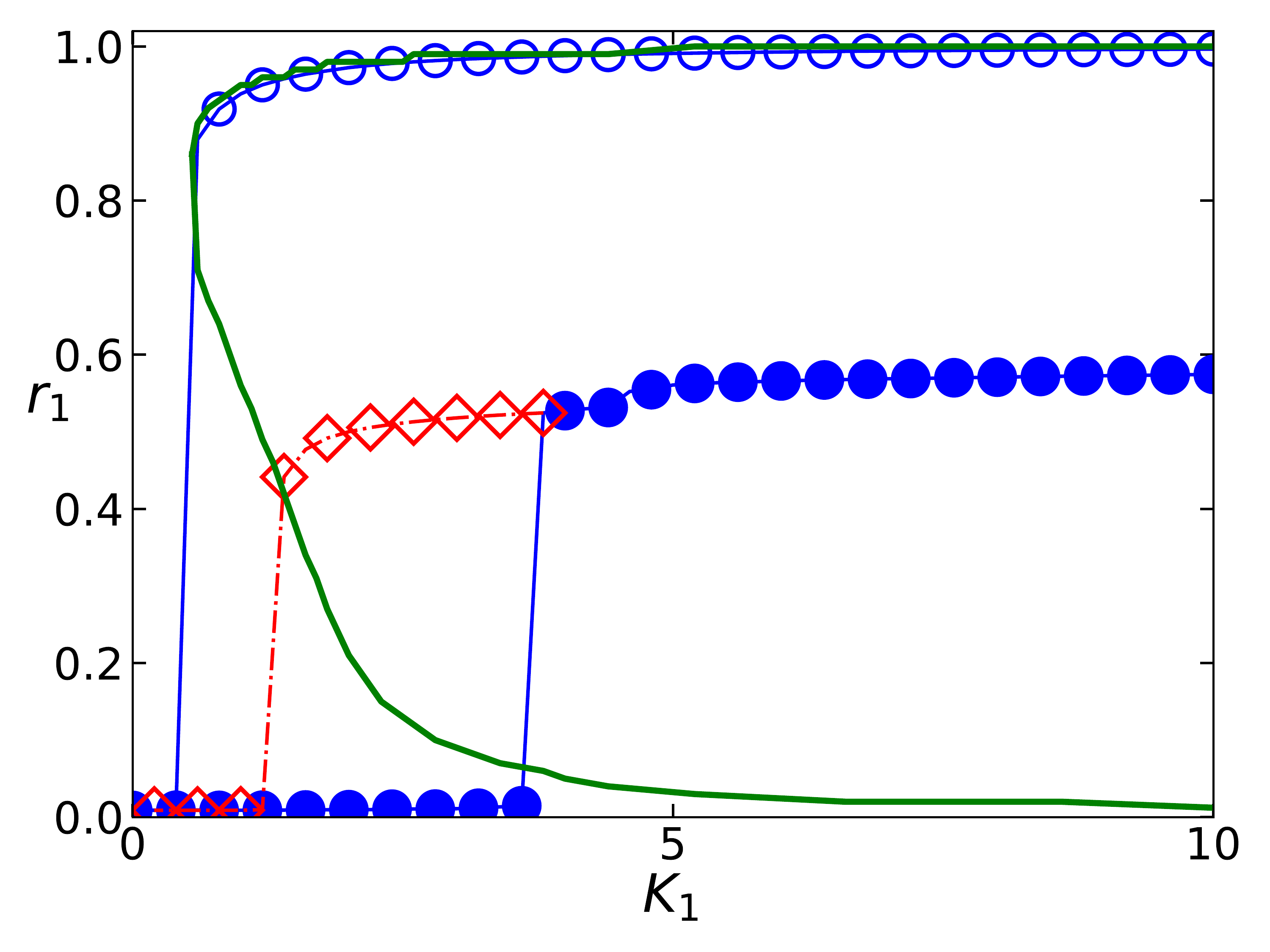}
    \caption{(Color online). Analytical results matching with numerics: Average $r_1$ vs $K_1$ for $N=10,000$. The analytical solution is denoted with green line, while the numerical results are shown as the blue solid (hollow) circles for forward (backward) process. Backward continuation from a multistable state denoted as red dash-dotted line (hollow diamond symbols) collides with the unstable branch, indicating saddle-node bifurcation. Here, $m=4, K_2 = 6$, and $\alpha=\pi/6$  
    }
    \label{fig:analytical_plot}
\end{figure}

Separating the real and imaginary part of the solution $z_2$, we get
\begin{multline}\label{a_r2}
    r_2  = \int_{-\infty}^{\infty}\Bigg[\mathbbm{1}_l((1-2\Gamma^2)\cos{2\beta} + 2\Gamma\sqrt{1-\Gamma^2} \sin{2\beta}) + \\ \mathbbm{1}_d M^2 (D\cos{2\beta} - 2 \Gamma\sin{2\beta}) \Bigg]  g(\omega) d\omega,
\end{multline}

\begin{multline}\label{a_psi2}
    0  = \int_{-\infty}^{\infty}\Bigg[\mathbbm{1}_l(2\Gamma\sqrt{1-\Gamma^2} \cos{2\beta} - (1-2\Gamma^2) \sin{2\beta}) - \\ \mathbbm{1}_d M^2 (D\sin{2\beta}+2 \Gamma \cos{2\beta}) \Bigg]  g(\omega) d\omega,
\end{multline}

where, $ M= \left(\sqrt{\frac{\Gamma^2}{\rho^2}-\frac{\rho^2}{\Gamma^2+\rho^4}} - \frac{\Gamma}{\rho}\right)$ and $D =\left( \frac{\Gamma^2}{\rho^2}- \rho^2 \right)$, while $\mathbbm{1}_l$ and  $\mathbbm{1}_d$ are indicator functions that correspond to the locked and drift case, respectively. Eq.~(\ref{a_r1} - \ref{a_psi2}) describes the steady-state behavior of the system in the backward process. Following Ref. \cite{gao2020}, one can approximate $\Omega$ as a function of the auxiliary parameter $R$ and $\beta$ for the case of the backward branch. By assuming that all oscillators are globally synchronized ($r_1 \approx 1$) at high $K_1$ limit, Eq.~(\ref{a_r1}-\ref{a_psi1}) can be rewritten as:
\begin{equation}\label{omR}
    \sin{\beta} = \int (\mathbbm{1}_l \Gamma - \mathbbm{1}_d M \rho)g(\omega) d\omega,
\end{equation}
and
\begin{equation*}
     \cos{\beta} = \int (\mathbbm{1}_l \sqrt{1-\Gamma^2} - \mathbbm{1}_d M (\Gamma/\rho))g(\omega) d\omega.
\end{equation*}

When the system achieves global coherence, Eq.~(\ref{omR}) contains only the locked contribution of oscillators and one gets
\begin{equation}\label{beta_eq}
    \sin{\beta} \approx \int_{-\omega_b +\Omega}^{\omega_b +\Omega} g(\omega) \left(\frac{\omega - \Omega}{R}\right) d\omega,
\end{equation}
where, $-\omega_b +\Omega \ll 0$ and $\omega_b +\Omega \gg 0$. Using normalization condition of $g(\omega)$, and noticing that for high $K_1$ value, $\beta \approx \alpha$, Eq.~(\ref{beta_eq}) follows
\begin{equation}\label{omega_eq}
    \Omega \approx -R \sin{\alpha}.
\end{equation}

As Eq.~(\ref{omega_eq}) is true only for the backward limit, the simulation results for Eq.~(\ref{a_r1} - \ref{a_psi2}) are shown in Fig.~(\ref{fig:analytical_plot}) for the backward branch showing excellent agreement at high $K_1$ value.

 In the following, we discuss the individual effects of phase lag, higher-order, and inertia on the system in detail:
\begin{figure}[t]  
    \centering
        \includegraphics[width=0.4\textwidth]{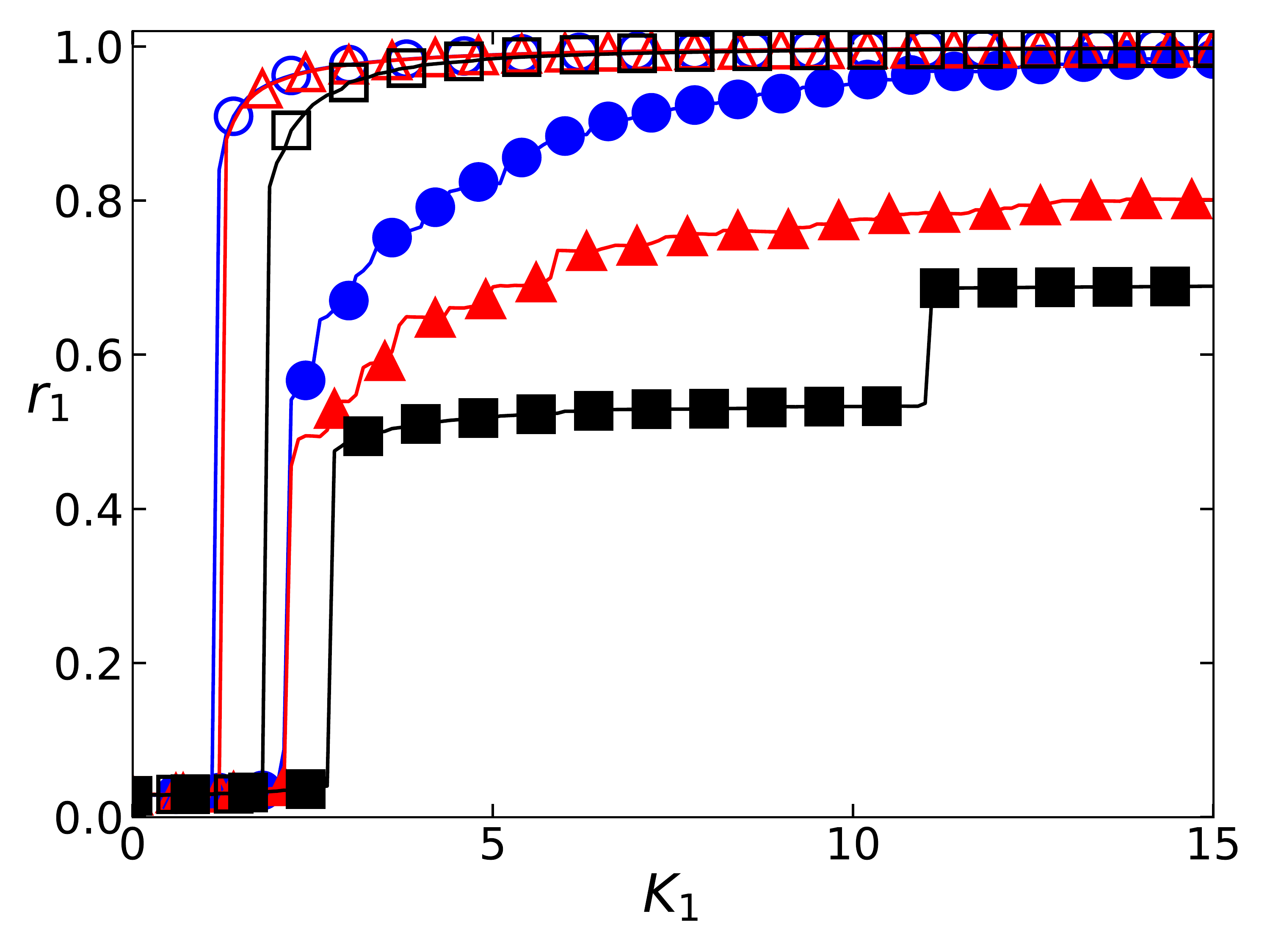}
    \caption{(Color online) Effect of phase lag. Average $r_1$ vs $K_1$ plot for $N=1000$ oscillators for different phase lag values.    The solid (hollow) symbols denote the forward (backward) process, while the shape of the symbols, i.e. blue circles, red triangles, black squares denote the different phase lag values $\alpha = 0.0, 0.1, \pi/6$, respectively. Here, $m=4$ and $K_2=2$.
    }
    \label{fig:running A}
\end{figure}

\subsection{Effect of phase lag \texorpdfstring{$\alpha$}:}
Fig.~(\ref{fig:running A}) shows the behavior of $r_1$ with $K_1$ following procedures-(I) and (II) with different $\alpha$ values for fixed $m$ and $K_2$. As the phase lag parameter causes frustration in the system, it delays synchronization; therefore, the forward transition point should increase with an increase in $\alpha$. As expected, from Fig.~(\ref{fig:running A}), the forward transition point associated with the procedure (I) $K_{1c}^f (\approx 2, 2.1, 2.7)$ increases with $\alpha (= 0.0, 0.1, \pi/6)$. A stepwise explosive synchronization appears in the forward direction for a higher value of phase lag (i.e., $\alpha=\pi/6$). It should also be noted that the backward transition point $K_{1c}^b (\approx 1.1, 1.2, 1.8)$ also increases with increasing value of phase lag parameter $\alpha (=0.0, 0.1, \pi/6)$.

The presence of a phase lag along with inertia opposes global synchronization in the forward process, and the associated hysteresis width is reduced compared to the case without phase lag. Note that for $\alpha =0$, the oscillators synchronize and exhibit clustering behavior for a sufficiently large coupling strength, leading to global coherence. However, introducing a phase lag strengthens the separation between clusters, and the order parameter exhibits large oscillations in the forward case.

\subsection{Effect of triadic interaction \texorpdfstring{$K_2$}:}
\begin{figure}[t]  
    \centering
        \includegraphics[width=0.4\textwidth]{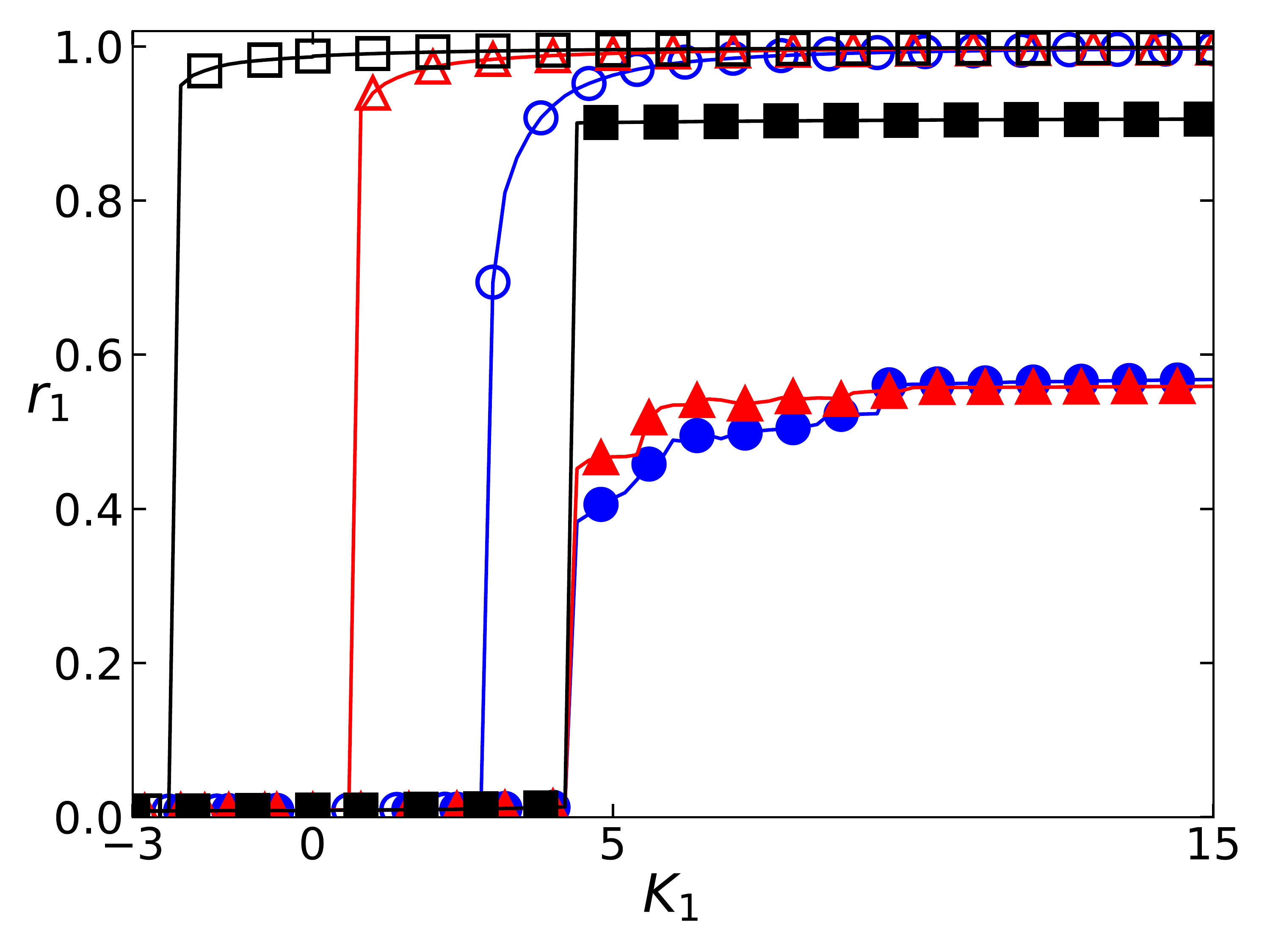}
    \caption{(Color online) Effect of $K_2$. Average $r_1$ as a function of $K_1$ for different value of $K_2$ for $N=10,000$ oscillators. The solid (hollow) symbols denote the forward (backward), while the shape of the symbols, i.e. blue circles, red triangles, black squares denote the different values of $K_2 = 0, 4, 8$, respectively. Here, $m=4$ and $\alpha= \pi/6$.
    }
    \label{fig:running K2}
\end{figure}

For a fixed value of $m=4$ and $\alpha=\pi/6$, Fig.~(\ref{fig:running K2}) shows the behavior of $r_1$ with $K_1$ for different values of $K_2$. The transient dynamics upto $t=2500$ is discarded, and the measured quantities were obtained by averaging over $t\in [2500,3900]$. Here, we take $N=10,000$ oscillators to minimize the finite-size effect \cite{ayushi}. Higher-order interactions only influence backward transition points and increase the height of jumps (Fig.~\ref{fig:running K2}) from incoherent to coherent state \cite{narayan2024}. These heights are defined as the difference in the value of $r_1$ between the partially coherent and incoherent states. A similar behavior of the forward transition point $K_{1c}^f (\approx 4.2)$ obtained by procedure (I) remains unchanged for different values of $K_2$. However, the backward transitions determined using procedure (II) correspond to $K_{1c}^b (\approx 2.8, 0.6, -2.3)$ for different values of $K_2 (=0.0,4.0,8.0)$. Note that the magnitude of the discontinuous jump from incoherent to coherent state in the forward case depends on the value of $K_2$ (Fig.~\ref{fig:running K2}). The magnitude of the jumps ranges from $r_1\approx 0$ to $r_1(\approx 0.4,0.45,0.9)$ for the corresponding set of $K_2 (=0.0,4.0,8.0)$ values, respectively. 

\subsection{Effect of inertia \texorpdfstring{$m$}:}
\begin{figure}[t]
    \centering
        \includegraphics[width=0.4\textwidth]{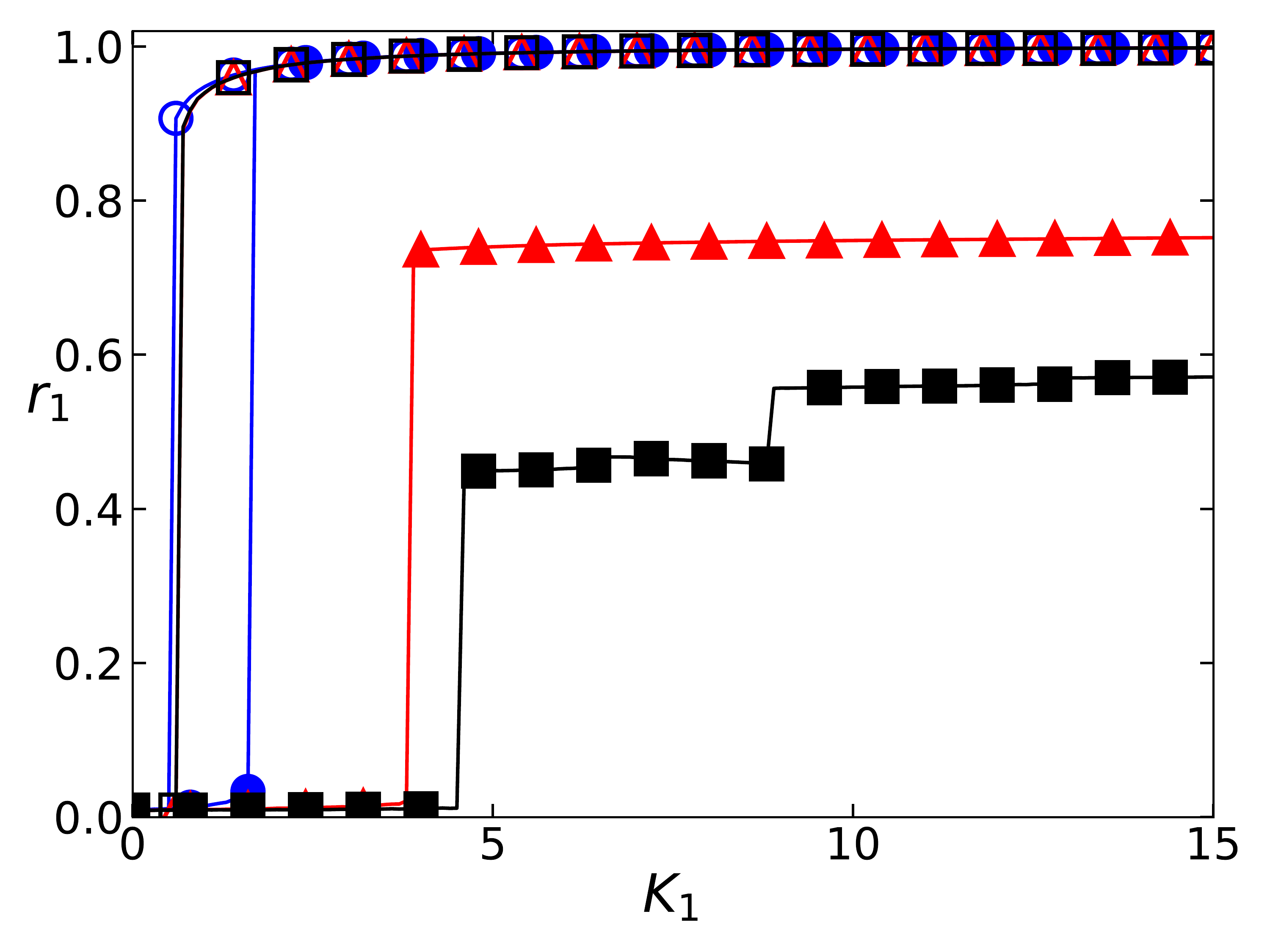}
    \caption{(Color online)Effect of $m$. Average $r_1$ as a function of $K_1$ for different value of $m$ for $N=10,000$ oscillators. The solid (hollow) symbols denote the forward (backward), while the shape of the symbols, i.e. blue circles, red triangles, black squares denote the different values of $m = 0, 2, 8$, respectively. Here, $K_2=4$ and $\alpha= \pi/6$}
    \label{fig:running m}
\end{figure}

Fig.~(\ref{fig:running m}) describes how the dynamics change with $m$. After discarding an initial transient, a time average of order parameter is measured. Again, to minimize finite-size effects, we consider a system of $10,000$. Upon increasing inertia $m(=0.0,2.0,8.0)$, the forward transition point associated with procedure (I) shifts to higher values $K_{1c}^f(\approx 1.6, 3.8, 4.5)$, while the backward transition associated with procedure (II) remains nearly the same, i.e. $K_{1c}^b(\approx 0.5,0.6,0.6)$. Additionally, the magnitude of the first-order jump $r_1 (\approx 0.95, 0.78,0.42)$ associated with the transition from incoherent to coherent state decreases with increasing $m$.

\begin{figure}[t]
    \centering
    \begin{tabular}{cc}

        \includegraphics[width=0.23\textwidth]{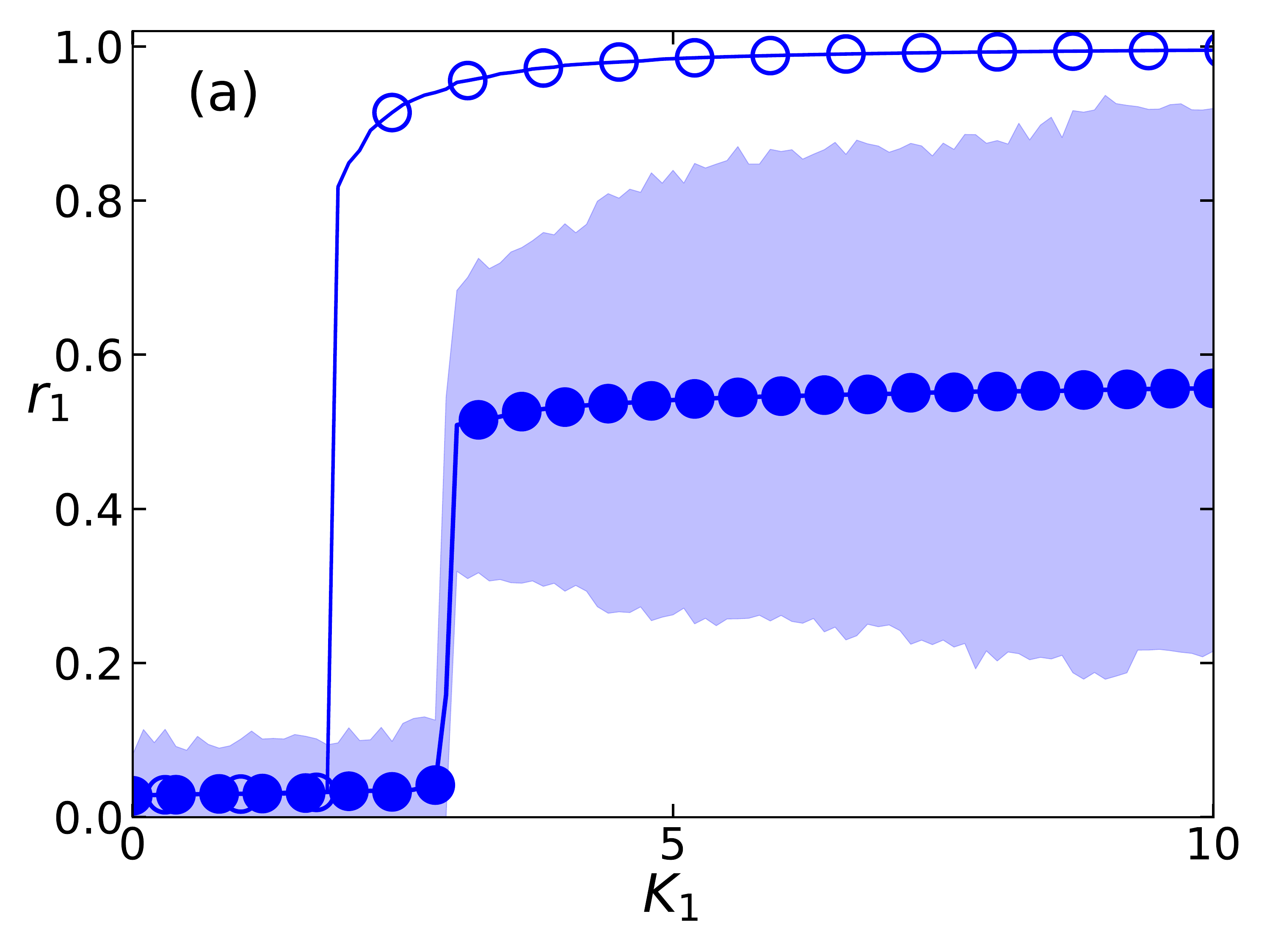} &
        \includegraphics[width=0.23\textwidth]{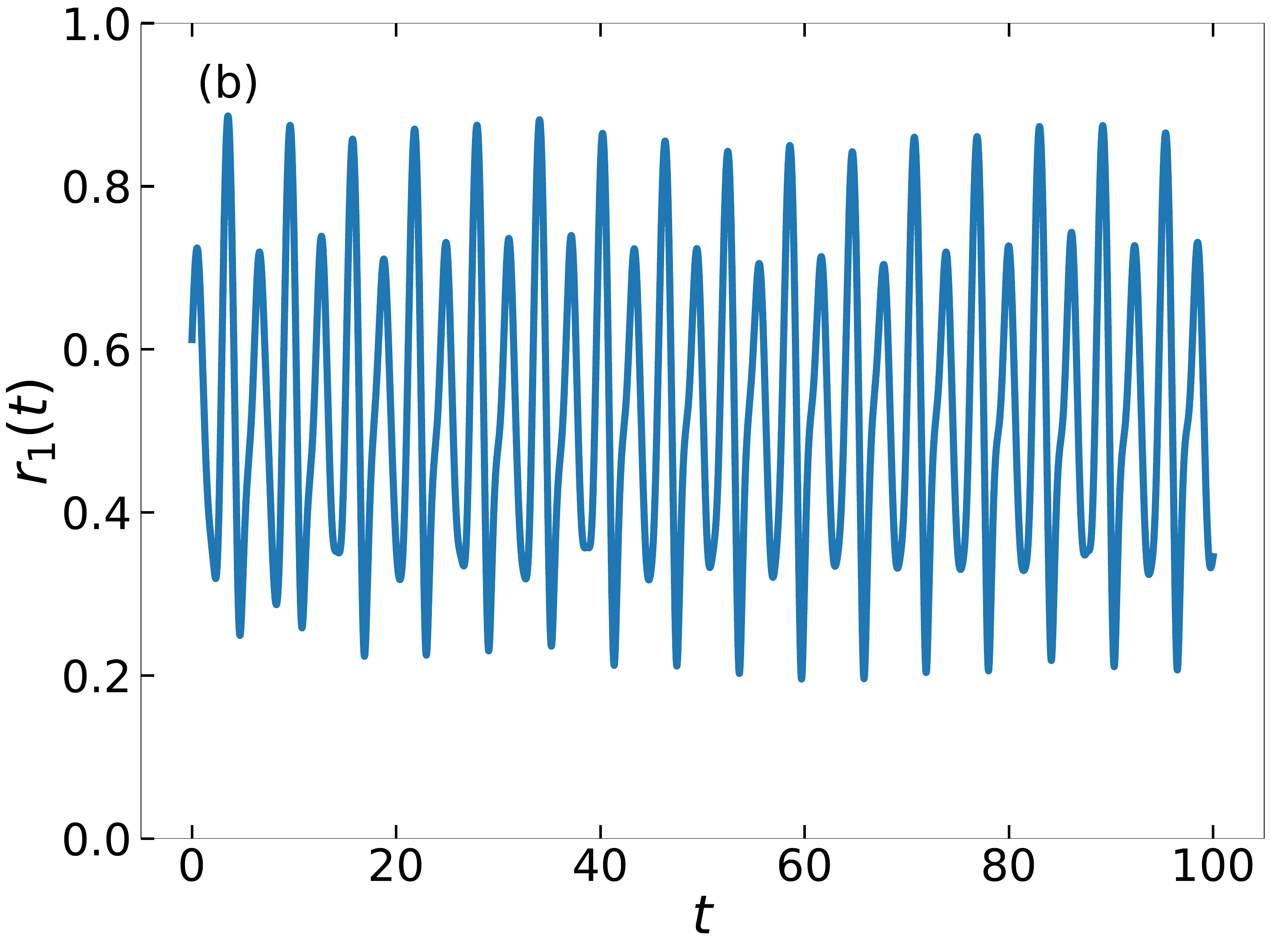} \\

        \includegraphics[width=0.23\textwidth]{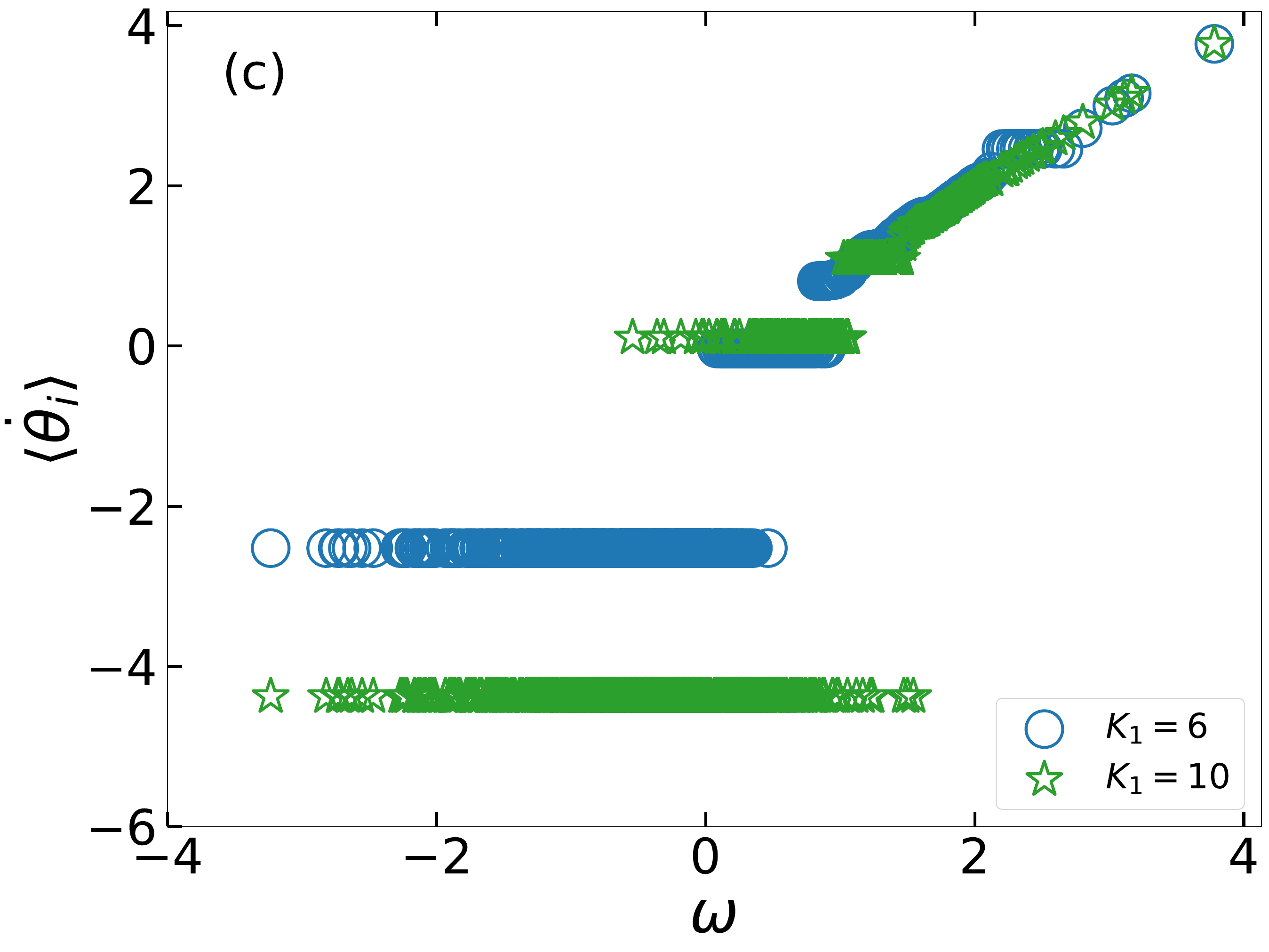} &
        \includegraphics[width=0.23\textwidth]{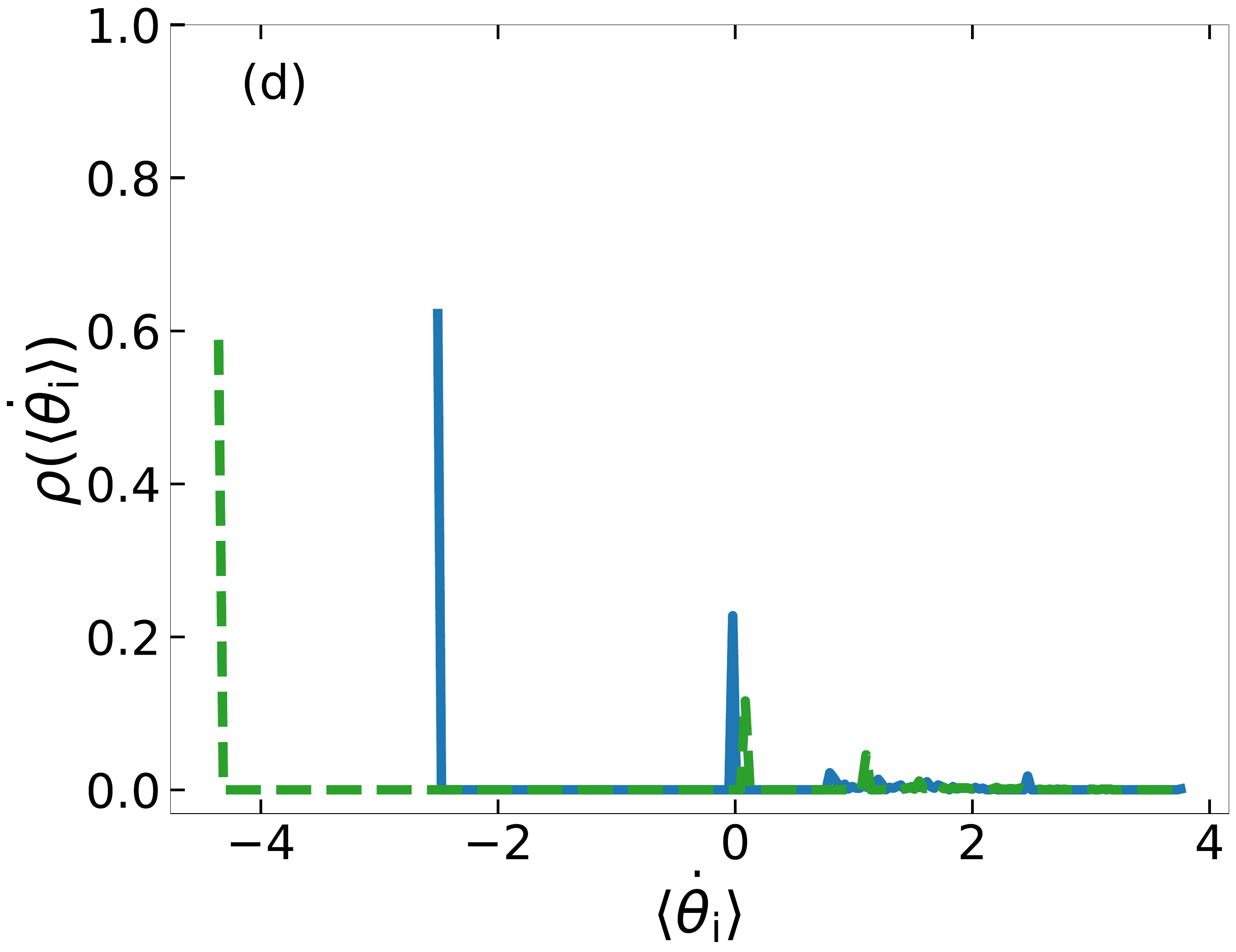} \\

    \end{tabular}

    \caption{(Color online) Numerical simulation for $N=1000$ oscillators with fixed $m=4$, $K_2=2$, and $\alpha=\pi/6$. (a) $r_1$ as a function of $K_1$ where blue solid (hollow) circles denote forward (backward) branches. The oscillation width is computed from $(r_1)_{\max}$ and $(r_1)_{\min}$. (b) The oscillation of $r_1$ as a function of time $t$ for $K_1=6$. (c) Average phase velocity $\langle\dot{\theta}_i\rangle$ vs natural frequency $\omega$ for $K_1=6$ (blue circles) and $K_1=10$ (green stars). (d) Density $\rho(\langle\dot{\theta}_i\rangle)$ vs $\langle\dot{\theta}_i\rangle$ showing clustering behavior for $K_1=6$ (blue solid line) and $K_1=10$ (green dashed line).}
    \label{fig:cluster_sync_1}

\end{figure}

As shown in Fig.~(\ref{fig:multistable})-~(\ref{fig:running m}) oscillators cannot achieve global synchronization when $\alpha$ and $m$ both have high values. The emergence of stepwise structure at sufficiently high values of $\alpha$ and $m$ in the forward process is due to the breakdown of the independence of drifting oscillators and the formation of locked oscillators at nonzero velocities. Fig.~(\ref{fig:cluster_sync_1}-a) depicts the behavior of $r_1$ with respect to time $t$ for $1000$ oscillators after discarding the transient from $t\in [0,4000]$. The value of $r_1$ does not remain constant, but oscillates around a mean value.

In Fig.~(\ref{fig:cluster_sync_1}-a), the mean value is a thick blue line, while the width of the oscillation is presented by the light blue band around the mean value. The states shown in Fig.~(\ref{fig:cluster_sync_1}-b) referred to as oscillating states\cite{olmi2015,gao2021} around the mean value of $r_1$ is shown over the interval $t\in [4000,4100]$ sampled at every $\delta t =0.01$ for $K_1=6$. Fig.~{(\ref{fig:cluster_sync_1}}-c) shows the average phase velocities of oscillators $\langle\dot{\theta}_i\rangle$ vs natural frequency $\omega_i$ for two different $K_1$ values, showing a shift in the average phase velocity ($\langle\dot{\theta}_i\rangle$) as a function of $K_1$ for fixed $m$ and $K_2$. A group of oscillators with the same value of average phase velocity forms synchronized (or \textit{frequency locked}) clusters. It is also clear that with an increase in $K_1$, more oscillators gets locked and the clusters become more prominent. Fig.~(\ref{fig:cluster_sync_1}-d) shows the density of average phase velocities $\rho(\langle\dot{\theta}_i\rangle)$ as a function of $\langle\dot{\theta}_i\rangle$ for two different $K_1$ values revealing two prominent clusters and their separation increases  with an increase of $K_1$.  

Next, we demonstrate that, due to the cross effects of $m$ and $\alpha$, one cannot achieve global synchronization in the forward process by assuming two-cluster behavior.

\section{Two-cluster synchronization}\label{two-cluster}
We quantify the model Eq.~(\ref{model}) with a two-cluster approximation and show that phase lag and inertia influence the system in such a way that it cannot achieve global synchronization even for arbitrary large coupling. 

For simplicity, we assume the existence of only two clusters, with the number of oscillators distributed among two distinct natural frequencies, i.e. $N_1$ oscillators with $\omega_1$ and $N_2$ oscillators with $\omega_2$. Hence, Eq.~(\ref{model}) can be simplified as follows:
\begin{equation}\label{theta1}
    m \ddot{\theta}_i + \dot{\theta}_i = \omega_1 + H(\theta_i,\alpha),
\end{equation}
and
\begin{equation}\label{theta2}
     m \ddot{\theta}_i + \dot{\theta}_i = \omega_2 + H(\theta_i, \alpha),
\end{equation}
where, $i=1,2,....,N_1$ for Eq.~(\ref{theta1} and $i=N_1+1,N_1+2,...,N$ for Eq.~(\ref{theta2} with the condition of $N=N_1+N_2$. The function $H(\theta_i, \alpha)$ is given as follows, 
\begin{multline}\label{ht}
    H(\theta_i,\alpha)=\frac{K_1}{N} \sum_{j=1}^N \sin(\theta_j - \theta_i - \alpha) \\+ \frac{K_2}{N^2}\sum_{j,k=1}^N \sin(2\theta_j - \theta_k - \theta_i - \alpha).
\end{multline}
These oscillators are split into two groups and are synchronized within each group. We define $\theta_1$ and $\theta_2$ as common phases in these two groups, respectively, and the fraction of oscillators within each group as $n_p = N_p/N$, where $p=1,2$. We remark that the normalization condition requires $n_1+n_2 = 1$.
Introducing a new variable $\xi= \theta_1-\theta_2$ and assuming $n_2>n_1$ without loss of generality, Eq.~(\ref{theta1}) and ~(\ref{theta2}) combined with Eq.~(\ref{ht}) gives as follows:
\begin{equation}\label{xi_eq}
    m\ddot{\xi} + \dot{\xi} = \Delta\Omega - a \sin{(\xi + \eta)} -b \sin{2\xi}
\end{equation} 
where,  $\Delta n = (n_2-n_1)$;
\begin{align*}
    \Delta\Omega &= (\omega_1-\omega_2) + (K_1+K_2)\Delta n\sin{\alpha}; \\
    a &= \sqrt{(K_1+K_2\Delta n^2)^2\cos^2{\alpha} + (K_1+K_2)^2\Delta n^2 \sin^2{\alpha}};\\
    \eta &= \arctan{\left(\frac{(K_1+K_2)\Delta n \tan{\alpha}}{K_1+K_2\Delta n^2}\right)} ;\hspace{0.3cm}
    b = 2 K_2 n_1n_2\\  
\end{align*}
Due to the transcendental nature of Eq.~(\ref{xi_eq}), it is difficult to analyze the full parameter space. However, our motivation is to show how the mixture of inertia and phase lag opposes global synchronization. As shown in Fig.~(\ref{fig:running K2}), forward transition point $K_{1c}^f$ is independent of $K_2$, and so it will be more useful to study Eq.~(\ref{xi_eq}) for the limit $K_2 \rightarrow0$ as studied by Gao and Efstathiou\cite{gao2020}. 

In the limit of $K_2 \rightarrow 0$, the synchronization condition for Eq.~(\ref{xi_eq}) is determined by two parameters\cite{gao2020}:
\begin{equation}
    \mathcal{A} = \frac{1}{\sqrt{mK_1q(\alpha)}}, \hspace{0.3cm} \mathcal{B} = \frac{\Delta \Omega}{K_1 q(\alpha)} = \frac{\omega_1-\omega_2}{K_1 q(\alpha)} + \frac{\Delta n \sin{\alpha}}{q(\alpha)}.
\end{equation}
where, 
\begin{align*}
    \Delta\Omega &= (\omega_1-\omega_2) + K_1\Delta n\sin{\alpha} \\
    a &=  K_1\sqrt{\cos^2{\alpha} + \Delta n^2 \sin^2{\alpha}} = K_1 q(\alpha)\\
    \eta &= \arcsin{\left(\frac{\Delta n \sin{\alpha}}{q(\alpha)}\right)};\hspace{0.5cm} b=0
\end{align*}
The condition for the oscillator to be in a fixed point region is $\mathcal{B}\leq \mathcal{B}_L(\mathcal{A})= (\Delta \Omega/K_1q(\alpha)) = 1$ and the limit cycle is given as
\begin{equation*}
    \mathcal{B} \geq \mathcal{B}_S(\mathcal{A}) =\begin{cases}
(4/\pi) \mathcal{A} - 0.305 \mathcal{A}^3, & \text{if } \mathcal{A} \leq 1.193, \\[2mm]
1, & \text{if } \mathcal{A} > 1.193.
\end{cases}
\end{equation*}

For $\omega_1 = \omega_2$ case, As $K_1 \rightarrow \infty$, we have $\mathcal{A} \rightarrow 0$ and $\mathcal{B} \rightarrow \Delta n \sin{\alpha}/q(\alpha) > 0$. From the fact that $\mathcal{B}_S(\mathcal{A}) \rightarrow 0$ in the limit of $K_1 \rightarrow \infty$, the condition for global synchronization $|\mathcal{B}| \leq \mathcal{B}_S(\mathcal{A})$ cannot be satisfied in the forward path followed as procedure (I). Hence, we do not have global synchronization. It is important to note that as $K_1 \rightarrow \infty$, $\mathcal{B}_L \rightarrow 0$, but $|\mathcal{B}|> 0 $ due to phase lag $\alpha$. Hence, the condition for the oscillator to be in the fixed point region is not satisfied. We conclude that in the forward process, the synchronized clusters at high $K_1$ values are formed partially due to drifting oscillators, which leads to $\langle \dot{\theta}_i \rangle$ as a function of $K_1$.

For the case of $\alpha \rightarrow 0$, the global synchronization condition is satisfied\cite{narayan2024} and one can get global synchronization. For the case of $m \rightarrow 0$, as $\mathcal{B}_S (\mathcal{A}) = 1$, while $\Delta n \sin{\alpha}/q(\alpha) <1$, we have $|\mathcal{B}|<1=\mathcal{B}_S$ in the limit of $K_1 \rightarrow \infty$. As a result, we can also achieve global synchronization for the case of $m \rightarrow 0$. It is clear that the partially synchronized behavior arises due to the combined effect of $m$ and $\alpha$.

\section{Conclusion}\label{conclusion}
In this study, we examined the impact of phase lag on the Kuramoto model with inertia, accounting for both pairwise and higher-order interactions. We showed that the phase lag term significantly alters the dynamics of the model, and the system exhibits distinct hysteresis widths, depending upon the chosen multistable branch as initial condition. This phenomena originates from the fact that the saddle node bifurcation occurs at different coupling strength value.

We systematically examined the individual role of phase lag, triadic interactions, and inertia on overall dynamical behavior of the system. The phase lag acts as a frustration in the system, and shifts both transition points towards higher coupling values. The inclusion of higher-order term enhances the bistability region by lowering the value of the backward transition point. The inertia term plays no visible role in the backward transition point, and decreases the height of the forward jumps with increasing inertia. We demonstrate that for the forward branch, phase lag tries to oppose global synchronization. As $K_1$ varies, the average phase velocity of the synchronized cluster changes, primarily due to the contribution of drifting oscillators. The formation of partially synchronized states is due to the cross-effect of inertia and phase lag leading to oscillatory states. 

A straightforward extension of the present work is to incorporate different phase lag values in pairwise and triadic terms.
The phase lag in pairwise coupling influences the forward transition point, whereas the phase lag associated with the triadic interactions governs the backward transition point. The width of the hysteresis can be tuned by changing phase lag values in both pairwise and triadic interactions.

The multiple hysteresis width associated with different multistable branch can be useful to model for selecting and storing the memory of the states. It is suggested \cite{skardal,xu} that higher-order is important for information storage without fine-tuning. The model also has potential applications in power grid systems and Josephson junction arrays \cite{app1}.

\appendix
\section{Calculation of $z_1$ and $z_2$ with the relation of $\psi_2$ and $\psi_1$}\label{app1}
For locked oscillators, 
\begin{equation}
    z_p^{(l)} = \int_{-\omega_{f,b} +\Omega} ^{\omega_{f,b} +\Omega}(\cos{p\theta^*} + i \sin{p\theta^*}) g(\omega) d\omega
\end{equation}
Putting $\theta^* = \arcsin{(\Gamma)} - \beta$, we get
\begin{multline}
    z_1 ^{(l)} = \int_{-\omega_{f,b} +\Omega}^{\omega_{f,b} + \Omega}(\sqrt{1-\Gamma^2}\cos{\beta} + \Gamma \sin{\beta}) + \\ i (\Gamma \cos{\beta} - \sqrt{1-\Gamma^2} \sin{\beta}) g(\omega) d\omega
\end{multline}
and 
\begin{multline}
    z_2 ^{(l)} = \int_{-\omega_{f,b}}^{\omega_{f,b}}((1-2\Gamma^2)\cos{2\beta} + 2\Gamma\sqrt{1-\Gamma^2} \sin{2\beta}) +\\ i (2\Gamma\sqrt{1-\Gamma^2} \cos{2\beta} - (1-2\Gamma^2) \sin{2\beta}) g(\omega) d\omega
\end{multline}

While, for drifting oscillators, We follow the standard method\cite{gao2018,narayan2024} where the $\dot{\theta}$ is written as the first harmonics of Fourier series as $[\dot{\theta} (\theta) = A_0 + A_1 \cos{\theta} + B_1 \sin{\theta}]$. On substituting Eq.~(\ref{red_eq}), we get

\begin{equation*}
    \dot{\theta}(\theta) = \nu_0 + \epsilon \cos{(\theta+\phi+\beta)}
\end{equation*}
where, $\nu_0 = \frac{\Gamma}{\rho}$, $\epsilon = \frac{\rho}{\sqrt{\rho^4+\Gamma^2}}$ and $\phi = \arcsin{\left(\frac{\rho^2}{\sqrt{\Gamma^2+\rho^4}}\right)}.$
\begin{figure}[t]
    \centering   \includegraphics[width=0.4\textwidth]{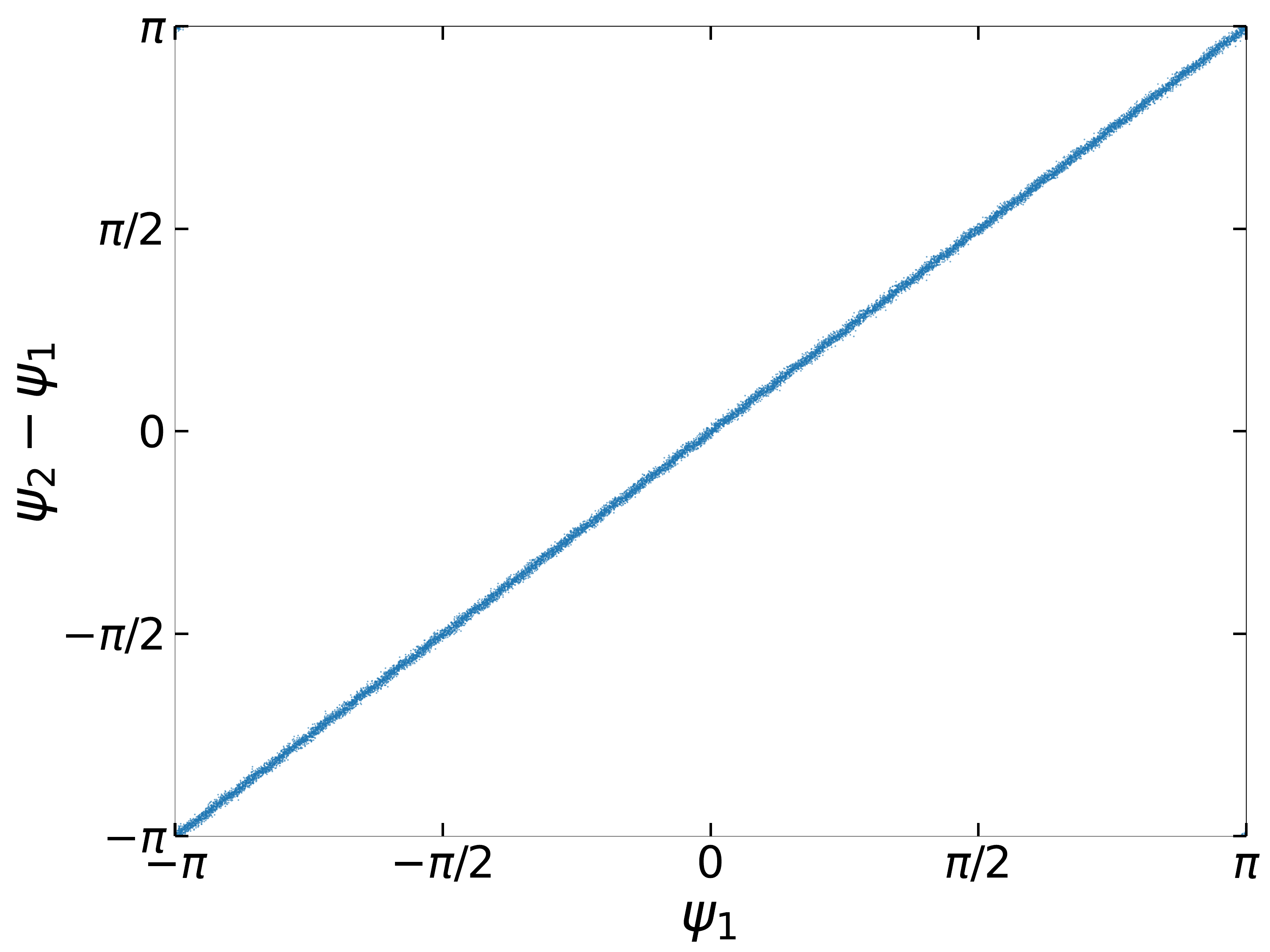}
    \caption{(Color Online) The relation between $\psi_2$ and $\psi_1$ for the backward process assuming the complete synchronization. Here, $K_1 = 15,K_2=6$, $m=4$ and $\alpha=\pi/6$.}
    \label{fig:psi2_psi1_relation}
\end{figure}
Putting this into Eq.~(\ref{ww}), we get 
\begin{multline}\label{A4}
     \langle e^{i p \theta} \rangle =\left[\frac{1}{T} \int_0^Te^{ip\theta} dt \right] \\ =e^{-ip\beta} (\nu_0 -i \rho)^p \big[-\nu_0 + \sqrt{\nu_0^2 - \epsilon^2}\big]^p
\end{multline}
and 
\begin{equation}\label{A5}
    z_p^{(d)} = \int_{|\omega-\Omega| > \omega_{f,b}} \langle e^{i p \theta} \rangle  g(\omega) d\omega
\end{equation}
Putting Eq.~(\ref{A4}) into Eq.~(\ref{A5}), we get
\begin{multline}
    z_1 ^{(d)} = \int_{|\omega|>\omega_{f,b}} M \Bigg[\Big(\frac{\Gamma}{\rho}\cos{\beta} - \rho\sin{\beta}\Big) \\- i \Big(\frac{\Gamma}{\rho}\sin{\beta}+ \rho \cos{\beta}\Big) \Bigg] g(\omega) d\omega
\end{multline}
and
\begin{multline}
    z_2 ^{(d)} = \int_{|\omega|>\omega_{f,b}}M^2 \Big
    ((D\cos{2\beta} - 2 \Gamma\sin{2\beta}) \\ -i (D\sin{2\beta}+2 \Gamma \cos{2\beta}))\Big] g(\omega) d\omega
\end{multline}

The dynamics corresponding to $z_1$ and $z_2$ can be simplified for the backward process assuming global synchronization. The relation between $\psi_1$ and $\psi_2$ is shown in Fig.~(\ref{fig:psi2_psi1_relation}) which implies $\psi_2 = 2 \psi_1$ giving $\gamma = 0$.
In a rotating frame, this implies that, $r_p = z_p^{(l)} + z_p^{(d)}$.

\bibliographystyle{apsrev4-2}
\bibliography{references}

@article{Winfree_book,
    author = {A. T. Winfree},
    title = {The Geometry of Biological Time},
    journal = {Springer},
    year = {2013}
}

@InProceedings{strogatz_book,
author="Strogatz, Steven H.",
editor="Levin, Simon A.",
title="Norbert Wiener's Brain Waves",
booktitle="Frontiers in Mathematical Biology",
year="1994",
publisher="Springer Berlin Heidelberg",
address="Berlin, Heidelberg",
pages="122--138",
isbn="978-3-642-50124-1"
}

@article{Strogatz20001,
title = {From Kuramoto to Crawford: exploring the onset of synchronization in populations of coupled oscillators},
journal = {Physica D: Nonlinear Phenomena},
volume = {143},
number = {1},
pages = {1-20},
year = {2000},
issn = {0167-2789},
author = {Steven H. Strogatz}
}

@book{osipov2007,
  title={Synchronization in Oscillatory Networks},
  author={Osipov, G.V. and Kurths, J. and Zhou, C.},
  isbn={9783540712688},
  lccn={2007927818},
  series={Springer Series in Synergetics},
  year={2007},
  publisher={Springer Berlin Heidelberg}
}

@article{Winfree1967,
title = {Biological rhythms and the behavior of populations of coupled oscillators},
journal = {Journal of Theoretical Biology},
volume = {16},
number = {1},
pages = {15-42},
year = {1967},
issn = {0022-5193},
author = {Arthur T. Winfree}
}

@InProceedings{kuramoto01,
author="Kuramoto, Yoshiki",
editor="Araki, Huzihiro",
title="Self-entrainment of a population of coupled non-linear oscillators",
booktitle="International Symposium on Mathematical Problems in Theoretical Physics",
year="1975",
publisher="Springer Berlin Heidelberg",
address="Berlin, Heidelberg",
pages="420--422",
isbn="978-3-540-37509-8"
}

@article{acebron2005,
  title = {The Kuramoto model: A simple paradigm for synchronization phenomena},
  author = {Acebr\'on, Juan A. and Bonilla, L. L. and P\'erez Vicente, Conrad J. and Ritort, F\'elix and Spigler, Renato},
  journal = {Rev. Mod. Phys.},
  volume = {77},
  issue = {1},
  pages = {137--185},
  numpages = {0},
  year = {2005},
  month = {Apr},
  publisher = {American Physical Society}
}

@article{networks,
    author = {Vasilyeva, E. and Kozlov, A. and Alfaro-Bittner, K. and Musatov, D. and Raigorodskii A. M. and Perc, M. and Boccaletti, S.},
    title = {Multilayer representation of collaboration networks with higher-order interactions},
    journal = {Scientific Reports},
    volume = {11},
    issue = {1},
    pages = {5666},
    year = {2021}
}

@article{RODRIGUES2016,
title = {The Kuramoto model in complex networks},
journal = {Physics Reports},
volume = {610},
pages = {1-98},
year = {2016},
issn = {0370-1573},
author = {Francisco A. Rodrigues and Thomas K. DM. Peron and Peng Ji and Jürgen Kurths}
}

@article{pikovsky_review,
    author = {Pikovsky, Arkady and Rosenblum, Michael},
    title = {Dynamics of globally coupled oscillators: Progress and perspectives},
    journal = {Chaos: An Interdisciplinary Journal of Nonlinear Science},
    volume = {25},
    number = {9},
    pages = {097616},
    year = {2015},
    month = {07}
}

@article{tanaka1997,
title = {Self-synchronization of coupled oscillators with hysteretic responses},
journal = {Physica D: Nonlinear Phenomena},
volume = {100},
number = {3},
pages = {279-300},
year = {1997},
issn = {0167-2789},
author = {Hisa-Aki Tanaka and Allan J. Lichtenberg and Shin'ichi Oishi}
}

@article{tanaka-prl,
  title = {First Order Phase Transition Resulting from Finite Inertia in Coupled Oscillator Systems},
  author = {Tanaka, Hisa-Aki and Lichtenberg, Allan J. and Oishi, Shin'ichi},
  journal = {Phys. Rev. Lett.},
  volume = {78},
  issue = {11},
  pages = {2104--2107},
  numpages = {0},
  year = {1997},
  month = {Mar},
  publisher = {American Physical Society}
}

@article{Ermentrout,
    author = {Ermentrout, B.},
    title = {An adaptive model for synchrony in the firefly Pteroptyx malaccae},
    journal = {Journal of Mathematical Biology},
    volume = {29},
    issue = {6},
    pages  = {571-585},
    year = {1991}
}

@article{levi1978,
 ISSN = {0033569X, 15524485},
 author = {M. Levi and F. C. Hoppensteadt and W. L. Miranker},
 journal = {Quarterly of Applied Mathematics},
 number = {2},
 pages = {167--198},
 publisher = {Brown University},
 title = {DYNAMICS OF THE JOSEPHSON JUNCTION},
 urldate = {2026-04-07},
 volume = {36},
 year = {1978}
}

@article{WATANABE1994,
title = {Constants of motion for superconducting Josephson arrays},
journal = {Physica D: Nonlinear Phenomena},
volume = {74},
number = {3},
pages = {197-253},
year = {1994},
issn = {0167-2789},
author = {Shinya Watanabe and Steven H. Strogatz}
}

@article{stroud2005,
  title = {Synchronization in disordered Josephson junction arrays: Small-world connections and the Kuramoto model},
  author = {Trees, B. R. and Saranathan, V. and Stroud, D.},
  journal = {Phys. Rev. E},
  volume = {71},
  issue = {1},
  pages = {016215},
  numpages = {20},
  year = {2005},
  month = {Jan},
  publisher = {American Physical Society}
}

@article{ikeda2011,
    author = {Y. Ikeda and H. Aoyama and Y. Fujiwara and H. Iyetomi and K. Ogimoto and W. Souma and H. Yoshikawa},
    title = {Coupled Oscillator Model of the Business Cycle with Fluctuating Goods Markets},
    journal = {arXiv 1110.6679},
    year = {2011}
}

@article{SAKYTE2011,
title = {Self-calming of a random network of dendritic neurons},
journal = {Neurocomputing},
volume = {74},
number = {18},
pages = {3912-3920},
year = {2011},
issn = {0925-2312},
author = {Edita Sakyte and Minvydas Ragulskis}
}

@article{nielsen2008,
    author = {Filatrella, G. and Nielsen, A. H. and Pedersen, N. F.},
    title = {Analysis of a power grid using a Kuramoto-like model},
    journal = {The European Physical Journal B},
    volume = {61},
number = {4},
pages = {485-491},
year = {2008}
}

@article{review,
title = {Universal nonlinear dynamics in damped and driven physical systems: From Pendula via Josephson junctions to power grids},
journal = {Physics Reports},
volume = {1147},
pages = {1-112},
year = {2025},
issn = {0370-1573},
author = {Subrata Ghosh and Linuo Xue and Arindam Mishra and Suman Saha and Dawid Dudkowski and Syamal K. Dana and Tomasz Kapitaniak and Jürgen Kurths and Peng Ji and Chittaranjan Hens}
}

@article{olmi2015,
  title = {Hysteretic transitions in the Kuramoto model with inertia},
  author = {Olmi, Simona and Navas, Adrian and Boccaletti, Stefano and Torcini, Alessandro},
  journal = {Phys. Rev. E},
  volume = {90},
  issue = {4},
  pages = {042905},
  numpages = {16},
  year = {2014},
  month = {Oct},
  publisher = {American Physical Society}
}

@article{gao2018,
  title = {Self-consistent method and steady states of second-order oscillators},
  author = {Gao, Jian and Efstathiou, Konstantinos},
  journal = {Phys. Rev. E},
  volume = {98},
  issue = {4},
  pages = {042201},
  numpages = {13},
  year = {2018},
  month = {Oct},
  publisher = {American Physical Society}
}

@article{p.ji2013,
author = {Ji, Peng and Peron, Thomas K. D. M. and Rodrigues, Francisco A. and Kurths, Jürgen},
title = {Low-dimensional behavior of Kuramoto model with inertia in complex networks},
journal = {Scientific Reports},
volume = {4},
SP = {4783},
year = {2014},
issn = {2045-2322}
}

@article{gao2021,
    author = {Gao, Jian and Efstathiou, Konstantinos},
    title = {Synchronized clusters in globally connected networks of second-order oscillators: Uncovering the role of inertia},
    journal = {Chaos: An Interdisciplinary Journal of Nonlinear Science},
    volume = {31},
    number = {9},
    pages = {093137},
    year = {2021},
    month = {09},
    issn = {1054-1500}
}

@article{sakaguchi1986,
    author = {Sakaguchi, Hidetsugu and Kuramoto, Yoshiki},
    title = {A Soluble Active Rotator Model Showing Phase Transitions via Mutual Entrainment},
    journal = {Progress of Theoretical Physics},
    volume = {76},
    number = {3},
    pages = {576-581},
    year = {1986},
    month = {09},
    issn = {0033-068X}
}

@article{olmi-chimera,
    author = {Olmi, Simona},
    title = {Chimera states in coupled Kuramoto oscillators with inertia},
    journal = {Chaos: An Interdisciplinary Journal of Nonlinear Science},
    volume = {25},
    number = {12},
    pages = {123125},
    year = {2015},
    month = {12},
    issn = {1054-1500}
}

@article{MUNYAYEV2024,
title = {Two-cluster regular states, chimeras and hyperchaos in a system of globally coupled phase oscillators with inertia},
journal = {Chaos, Solitons \& Fractals},
volume = {179},
pages = {114415},
year = {2024},
issn = {0960-0779},
author = {Vyacheslav O. Munyayev and Maxim I. Bolotov and Lev A. Smirnov and Grigory V. Osipov}
}

@article{cyclops,
  title = {Cyclops States in Repulsive Kuramoto Networks: The Role of Higher-Order Coupling},
  author = {Munyayev, Vyacheslav O. and Bolotov, Maxim I. and Smirnov, Lev A. and Osipov, Grigory V. and Belykh, Igor},
  journal = {Phys. Rev. Lett.},
  volume = {130},
  issue = {10},
  pages = {107201},
  numpages = {6},
  year = {2023},
  month = {Mar},
  publisher = {American Physical Society}
}

@article{ashwin, 
    author = {Ashwin, P., Christian B.},
    title = {Global Bifurcations Organizing Weak Chimeras in Three Symmetrically Coupled Kuramoto Oscillators with Inertia},
    journal = {Journal of Nonlinear Science},
    volume = {35},
    number = {45},
    year = {2025}
}

@article{krischer,
    author = {Schöhs,Yannick and Thomé, Nicolas and Krischer, Katharina},
    title = {Origin of Frequency Clusters and Robust Triplet Locking in the Kuramoto Model with Inertia},
    journal = {arXiv 2601.06595},
    year = {2026}
}

@article{barre,
  title = {Bifurcations and Singularities for Coupled Oscillators with Inertia and Frustration},
  author = {Barr\'e, J. and M\'etivier, D.},
  journal = {Phys. Rev. Lett.},
  volume = {117},
  issue = {21},
  pages = {214102},
  numpages = {5},
  year = {2016},
  month = {Nov},
  publisher = {American Physical Society}
}

@article{crawford1,
    author = {Crawford, John David},
    title = {Amplitude equations for electrostatic waves: Universal singular behavior in the limit of weak instability},
    journal = {Physics of Plasmas},
    volume = {2},
    number = {1},
    pages = {97-128},
    year = {1995},
    month = {01},
    issn = {1070-664X}
}

@article{crawford2,
  title = {Scaling and Singularities in the Entrainment of Globally Coupled Oscillators},
  author = {Crawford, John David},
  journal = {Phys. Rev. Lett.},
  volume = {74},
  issue = {21},
  pages = {4341--4344},
  numpages = {0},
  year = {1995},
  month = {May},
  publisher = {American Physical Society}
}

@article{gao2020,
    author = {Gao, Jian and Efstathiou, Konstantinos},
    title = {Synchronization of coupled second-order Kuramoto-Sakaguchi oscillators},
    journal = {arXiv 2012.14088},
    year = {2020}
}

@article{skardal,
  title = {Abrupt Desynchronization and Extensive Multistability in Globally Coupled Oscillator Simplexes},
  author = {Skardal, Per Sebastian and Arenas, Alex},
  journal = {Phys. Rev. Lett.},
  volume = {122},
  issue = {24},
  pages = {248301},
  numpages = {6},
  year = {2019},
  month = {Jun},
  publisher = {American Physical Society}
}

@article{xu,
  title = {Bifurcation analysis and structural stability of simplicial oscillator populations},
  author = {Xu, Can and Wang, Xuebin and Skardal, Per Sebastian},
  journal = {Phys. Rev. Res.},
  volume = {2},
  issue = {2},
  pages = {023281},
  numpages = {7},
  year = {2020},
  month = {Jun},
  publisher = {American Physical Society}
}

@article{priyanka1,
    author = {Rajwani, Priyanka and Suman, Ayushi and Jalan, Sarika},
    title = {Tiered synchronization in Kuramoto oscillators with adaptive higher-order interactions},
    journal = {Chaos: An Interdisciplinary Journal of Nonlinear Science},
    volume = {33},
    number = {6},
    pages = {061102},
    year = {2023},
    month = {06},
    issn = {1054-1500}
}

@article{narayan2024,
  title = {Prolonged hysteresis in the Kuramoto model with inertia and higher-order interactions},
  author = {Sabhahit, Narayan G. and Khurd, Akanksha S. and Jalan, Sarika},
  journal = {Phys. Rev. E},
  volume = {109},
  issue = {2},
  pages = {024212},
  numpages = {11},
  year = {2024},
  month = {Feb},
  publisher = {American Physical Society}
}

@article{OA,
    author = {Ott, Edward and Antonsen, Thomas M.},
    title = {Low dimensional behavior of large systems of globally coupled oscillators},
    journal = {Chaos: An Interdisciplinary Journal of Nonlinear Science},
    volume = {18},
    number = {3},
    pages = {037113},
    year = {2008},
    month = {09}
}

@InProceedings{Melnikov,
author="Guckenheimer, J. and Holmes, P.",
booktitle="Nonlinear Oscillations, Dynamical Systems, and Bifurcations of Vector Fields",
year="2013",
publisher="Springer Science \& Business Media, Berlin, 42.",
volume="42"
}

@article{ayushi,
    author = {Suman, Ayushi and Jalan, Sarika},
    title = {Finite-size effect in Kuramoto oscillators with higher-order interactions},
    journal = {Chaos: An Interdisciplinary Journal of Nonlinear Science},
    volume = {34},
    number = {10},
    pages = {101101},
    year = {2024},
    month = {10},
    issn = {1054-1500}
}

@article{app1,
  title = {Synchronization Transitions in a Disordered Josephson Series Array},
  author = {Wiesenfeld, Kurt and Colet, Pere and Strogatz, Steven H.},
  journal = {Phys. Rev. Lett.},
  volume = {76},
  issue = {3},
  pages = {404--407},
  numpages = {0},
  year = {1996},
  month = {Jan},
  publisher = {American Physical Society}
}

\end{document}